\documentclass[prb,reprint,amsmath,amssymb,showpacs,superscriptaddress,floatfix,longbibliography]{revtex4-1}
\usepackage[breaklinks=true,colorlinks,citecolor=blue,linkcolor=blue,urlcolor=blue]{hyperref}
\usepackage{float}
\usepackage{epsfig,mathrsfs,color,latexsym,subfigure,marginnote}

\newcommand{\nn}{\nonumber}

\newcommand{\bsigma}{\boldsymbol{\sigma}}
\DeclareMathAlphabet{\bi}{OML}{cmm}{b}{it}
\def\be{\begin{equation}}
\def\ee{\end{equation}}
\def\bearr{\begin{eqnarray}}
\def\eearr{\end{eqnarray}}

\def\bs{\boldsymbol}

\begin{document}
\title{Terahertz {shifted} optical sideband generation in graphene}
\author{Ashutosh Singh}
\affiliation{Department of Physics, Indian Institute of Technology Kanpur, Kanpur - 208016, India}
\author{Saikat Ghosh}
\affiliation{Department of Physics, Indian Institute of Technology Kanpur, Kanpur - 208016, India}
\author{Amit  Agarwal}
\email{amitag@iitk.ac.in}
\affiliation{Department of Physics, Indian Institute of Technology Kanpur, Kanpur - 208016, India}

\date{\today}

\begin{abstract}
Exploration of optical non-linear response of graphene predominantly relies on ultra-short time domain measurements. Here we propose an alternate technique 
that uses frequency {shifted} continuous wavefront optical fields, thereby probing graphene's steady state non-linear response. We predict  frequency sideband generation in the reflected field that originates from coherent electron dynamics of the photo-excited carriers. 
The corresponding threshold in input intensity for optimal sideband generation provides a direct measure of the third 
order optical non-linearity in graphene. Our formulation yields analytic forms for the generated sideband intensity, is applicable to generic two-band systems and suggests a range of applications that include switching of frequency sidebands using non-linear phase shifts and generation of frequency combs. 
\end{abstract}

\maketitle
 
\section{Introduction}
Owing to its gate tunable electronic, optical and opto-electronic properties, the exploration of {\it non-linear optical effects} in graphene has attracted significant interest in experiments \cite{Hafez2018,Yoshikawa2017,Prechtel2012,Jiang2018,Gu2012,Hendry,Yang,Zhang,Kumar,Hong} as well as in theory \cite{JPCC_Mikhailov,Glazov2014,Cheng2014,PhysRevB.90.245423,PhysRevLett.116.016601,PhysRevB.91.235320,PhysRevB.93.085403,PhysRevB.93.161411,Gullans2013,Yao2014}. Experimentally, several non-linear optical effects such as higher harmonic generation \cite{Yoshikawa2017,Hafez2018}, third order non-linearity \cite{Jiang2018,Prechtel2012,Hendry,Yang,Zhang,Kumar, Hong} and four wave mixing \cite{Gu2012} have been demonstrated in graphene. There are also predictions of ultra broad-band wave mixing at low powers, with 
generation of several side-bands at terahertz (THz) frequencies in bilayer graphene \cite{Crosse2014}. 
Such measurements and estimates offer fundamental insights in optical nonlinear interactions and relaxation mechanisms in lower dimensional systems \cite{Yoshikawa2017,Crosse2014,Prechtel2012,Cox2017,PhysRevLett.116.016601} along with a promise of applications including compact and useful THz sources and gate tunable opto-electronic devices \cite{Bonaccorso2010, Bao2012,PhysRevB.94.241107}. However, experiments till date have been primarily limited to ultra-short time-domain spectroscopy \cite{Dawlaty2008,PhysRevLett.111.027403} which are technologically involved and intricate, and physical interpretations generally rely on large scale computation.   

\begin{figure}[t!]
\includegraphics[width = 1.01\linewidth]{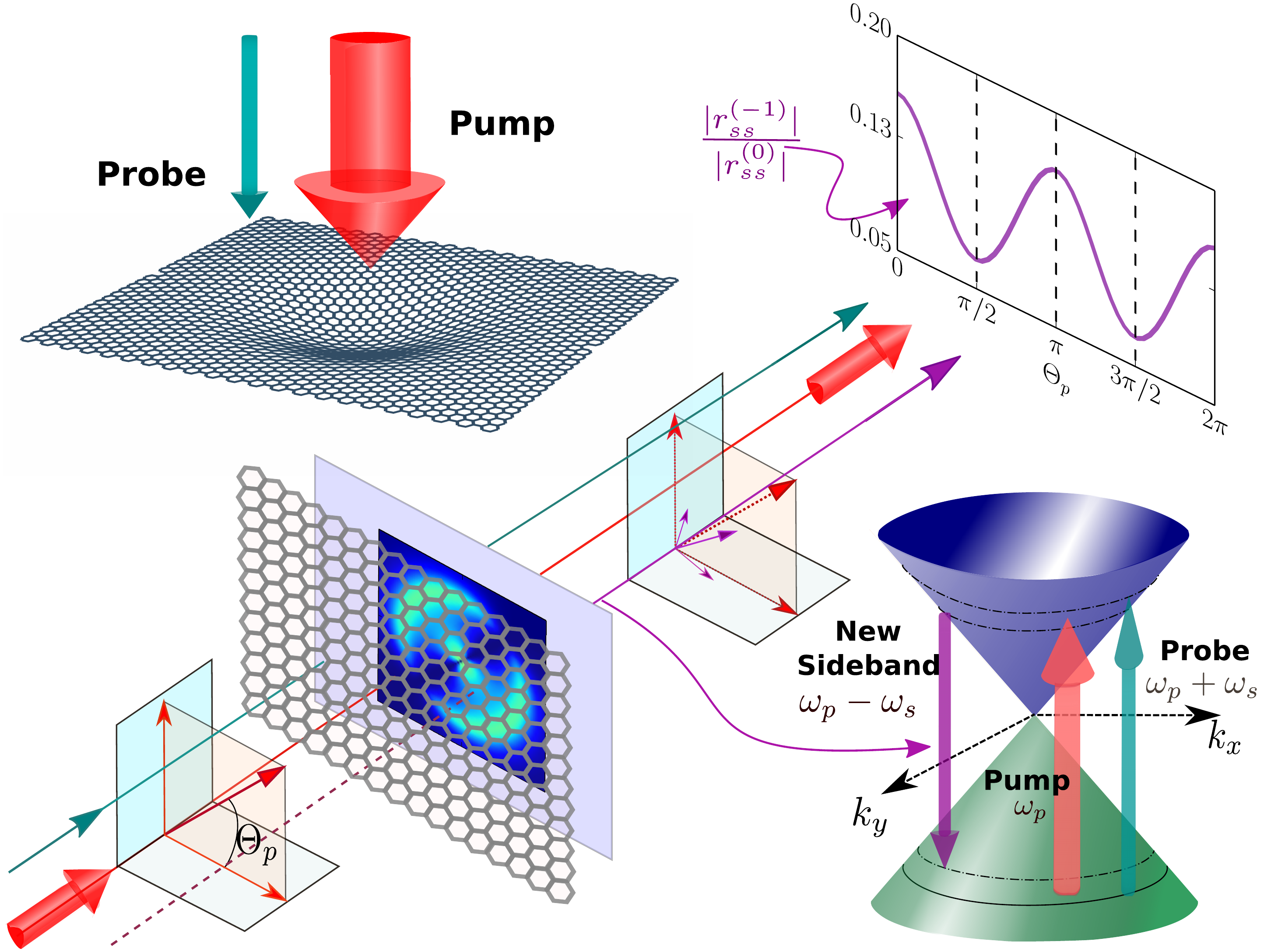}
\caption{Schematic of graphene illuminated by a CW pump beam (red) of frequency $\omega_p$ and a frequency {shifted} probe beam (green) of frequency $\omega_p +\omega_s$. The third harmonic related non-linear inter-band polarization combines with optical Bloch oscillations to generate 
a new side-beam (purple) at frequencies $\omega_p - \omega_s$. The color plot in the center shows the momentum resolved photo-excited population inversion corresponding to the pump beam: $n_{\bs k}^{(0)} = \rho_{cc}^{(0)} - \rho_{vv}^{(0)}$. 
The new sideband can be detected in the backward propagating beam via the optical intensity and polarization angle dependence of the reflectivity and polarization angle of the sideband. 
\label{Fig1}}
\end{figure}

Here we propose an alternate technique that uses frequency {shifted} continuous wavefront (CW) optical fields, probing optical non-linearity in the `steady state'.
In particular we focus on 
the non-linear optical sideband generation in graphene due to inter-band polarization combined with optical Bloch oscillations \cite{Boyd,Schubert2014}. In presence of a CW pump (frequency $\omega_p$) and a frequency {shifted} probe beam ($\omega_p + \omega_s$) the  optically pumped population inversion and the inter-band coherence oscillate at the modulation frequency. Such coherent `slushing' of the inter-band quasiparticles excitations leads to a new sideband generation at frequencies $\omega_p - \omega_s = 2 \omega_p - (\omega_p + \omega_s)$, as shown in Fig.~\ref{Fig1}. This results in distinct signatures in reflectivity along with non-linear polarization rotation at the new sideband frequency.  
Our formulation based on the dynamics of density matrix for a generic two band systems, can be easily applied to other materials as well. 

The predicted sideband generation is a direct consequence of non-degenerate four-wave mixing due to third-order non-linearity in graphene \cite{Boyd,Zhang1}. Estimation of the corresponding intensity threshold and polarization rotation offers an alternative technique for probing non-linear optical effects and relaxation rates with CW fields in graphene \cite{Zhang1}. Furthermore, the formulation is applicable from THz to optical domain with applications including switching with controlled non-linear phase shifts approaching $\pi/2$ with reasonable incident CW power and cascaded generation of frequency combs \cite{Burghoff2014}. 

\section{Two band model}
Our formulation starts with Hamiltonian of an electronic system interacting with an electro-magnetic field. It can be described using the dipole approximation \cite{Aversa}, i.e., 
$\hat H = \hat H_0 + e{\bf E}\cdot{\hat{\bf r}}$. Here $\hat H_0$ is the bare Hamiltonian, $e$ is the electronic charge, and ${\bf E}$ 
is the electric field. For simplicity we focus on a generic two band system \cite{PhysRevB.95.155421,PhysRevB.97.045402,PhysRevB.97.205420}, with its quasiparticle dispersion  described by the Hamiltonian, 
$
\hat H_0 = \sum_{\bs{k}}{\bf h}_{\bs k} \cdot \bsigma$,
where $ {\bf h}_{\bs k} = (h_{0{\bs k}},h_{1{\bs k}}, h_{2{\bs k}}, h_{3{\bs k}})$ is a vector composed of real scalar elements and $ \bsigma = (\openone_2, \sigma_x, \sigma_y, \sigma_z)$ is 
a vector composed of the $2 \times 2$ identity and the three Pauli matrices. The eigenvalues for conduction/valence band are, $\varepsilon^{c/v}_{\bs k} = h_{0{\bs k}} \pm g_{\bs k}$, where $g_{\bs k} \equiv \sqrt{h_{1{\bs k}}^2 + h_{2{\bs k}}^2 + h_{3{\bs k}}^2}$. 
Accordingly, the state vectors are given by, $\hat H_0|\psi^{c/v}_{\bs k}\rangle = \varepsilon^{c/v}_{\bs k}|\psi^{c/v}_{\bs k}\rangle$. 
The frequency {shifted}  electromagnetic field is, ${\bf E}(t) = {\rm Re}\left[{\bf E}_pe^{-i\omega_p t} + {\bf E}_se^{-i(\omega_p + \omega_s)t}\right],$
composed of a primary pump beam (of amplitude $|{\bf E}_p|$ and frequency $\omega_p$) and a probe beam (of amplitude $|{\bf E}_s| \ll |{\bf E}_p|$ and frequency $\omega_p+\omega_s$, where $\omega_s \ll \omega_p$). In general the pump and the probe fields can have different polarization angles, $\Theta_p$ and $\Theta_s$, respectively,  so that ${\bf E}_p = |{\bf E}_p|\left(\cos\Theta_p, \sin\Theta_p\right)$ and ${\bf E}_s = |{\bf E}_s|\left(\cos\Theta_s, \sin\Theta_s\right)$ for vertical incidence. 

The dynamics of the two band system described above is obtained by analytically solving the equation of motion (EOM) for the density matrix ($\rho^{ij}_{\bs k}$). The diagonal elements of $\rho_{\bs k}^{ij}$ comprise of the carrier distribution in the conduction ($\rho^{cc}_{\bs k}$) and valence ($\rho^{vv}_{\bs k}$) bands, while the off-diagonal elements  $\rho^{vc}_{\bs k} = (\rho^{cv}_{\bs k})^* \equiv p_{\bs k}$ capture the inter-band coherence. The incident optical field `pumps' the carriers from the valence band to the conduction band via vertical transitions. This optical pumping of carriers is countered by damping terms originating from the vacuum fluctuations, electron-electron interactions, electron-phonon interactions, and disorder, leading to a finite population inversion ($n_{\bs k} = \rho^{cc}_{\bs k}-\rho^{vv}_{\bs k}$) -- shown in Fig.~\ref{Fig1}. 
Including the damping terms phenomenologically in the EOM of the density matrix leads to the following set of coupled {\it optical Bloch equation} (OBE)~\cite{PhysRevB.94.195438,PhysRevB.95.155421, PhysRevB.97.045402, PhysRevB.97.205420},
\bearr\label{OBE_n}
\partial_t n_{\bs k} &=& 4~{\rm Im}\left[{\Omega}^{cv}_{\bs k}p_{\bs k}\right] - \gamma_1\left(n_{\bs k}-n^{\rm eq}_{\bs k}\right),\\
\partial_t p_{\bs k} &=& i\omega_{\bs k} p_{\bs k} - i{\Omega}^{vc}_{\bs k}n_{\bs k} - \gamma_2 p_{\bs k}\label{OBE_p}~. 
\eearr%
The inter-band Rabi frequency can be expressed in terms of the inter-band optical matrix element ${\bf M}^{vc}_{\bs k} \equiv \langle \psi^{v}_{\bs k}| e\nabla_{\bs k}\hat H_0/\hbar |\psi^{c}_{\bs k}\rangle$ as: 
$\hbar{\Omega}^{vc}_{\bs k} (t) = i{\bf E}(t)\cdot{\bf M}^{vc}_{\bs k}/\omega_{\bs k}$, where 
$\omega_{\bs k} = (\varepsilon^{c}_{\bs k} - \varepsilon^{v}_{\bs k})/\hbar$ is the vertical transition frequency 
\footnote{Strictly speaking, there is an additional term of the form $i (\Omega_{\bs k}^{cc} -\Omega_{\bs k}^{vv}) p_{\bs k}$ on the right hand side of Eq.~\eqref{OBE_p}. 
However this term can be safely neglected as it 1) oscillates with a frequency $~ 2\omega_p$ \cite{PhysRevB.95.155421} and 2) its impact on the final steady state dynamics turns out to be very small \cite{PhysRevB.97.205420}.}.
In Eqs.~\eqref{OBE_n}-\eqref{OBE_p}, $\gamma_1$ and $\gamma_2$ are the 
phenomenological damping rates of the inter-band population inversion and coherence, respectively. For simplicity, we assume these rates to be constants. 
{Note that this `constant damping rate' approximation allows us to proceed analytically, and it still captures all the relevant physics qualitatively. A more involved modelling of 
the damping constants, as done in Ref.~[\onlinecite{2018arXiv180610123S}], also yields a very similar results for the population inversion and the interband coherence.}

\section{Steady state solution}
For incident CW field, competition between optical pumping and decay rates, results in an eventual steady state. 
In this regime, analytical solutions can be obtained by making the following {\it ansatz} for the population inversion and inter-band coherence \cite{Boyd}:  
{\bearr \label{Eq:ansatz}
n_{\bs k} &=& n_{\bs k}^{(0)} + n_{\bs k}^{(1)} e^{-i\omega_s t} + n_{\bs k}^{(-1)} e^{i\omega_s t}~, \\  \label{Eq:ansatz2}
p_{\bs k} &=& \left[p_{\bs k}^{(0)} + p_{\bs k}^{(1)}e^{-i\omega_s t} + p_{\bs k}^{(-1)}e^{i\omega_s t}\right]e^{i \omega_p t}~. 
\eearr
Here, the superscript $(0)$ is used to denote the steady state solution of the OBEs in presence of CW pump field leading to optical response at frequency $\omega_p$ \cite{PhysRevB.94.195438,PhysRevB.95.155421, PhysRevB.97.045402, PhysRevB.97.205420}. 
{Presence of a probe field leads to slowly oscillating side-bands. Significant among these are the response (defined in terms of optical conductivities) at frequencies $ \omega_p \pm \omega_s$, denoted by superscript (1) and (-1) in Eq.~\eqref{sigma_1} and Eq.~\eqref{sigma_min_1} below. The side-band at $\omega_p - \omega_s$ is generated as a result of interaction between two photons of pump field with one photon of the probe field [$\omega_p - \omega_s = 2\omega_p - (\omega_p + \omega_s)$]. 
The {\it ansatz} in Eqs.~\eqref{Eq:ansatz}-\eqref{Eq:ansatz2} for the population inversion and inter-band coherence quantifies this. The addition of $n^{(-1)}_{\bs k} (=  [n^{(1)}_{\bs k}]^*)$, is mandated to make the population inversion real. This in turn forces the terms added in Eq.~\eqref{Eq:ansatz2} for the coherence which highlights the systems response at frequencies $\omega_p \pm \omega_s$.}

Using the {\it ansatz} of Eq.~\eqref{Eq:ansatz} in Eqs.~\eqref{OBE_n}-\eqref{OBE_p}, and taking long time average for the steady state, we obtain %
\be\label{n0}
\frac{n^{(0)}_{\bs k}}{n^{\rm eq}_{\bs k}} = \left(1 + \zeta^2\gamma_2^2
\frac{\omega^2}{\omega_{\bs k}^2}~\frac{|{\bf e}_p\cdot\tilde{\bf M}^{cv}_{\bs k}|^2}{(\omega_{\bs k} - \omega_p)^2 + \gamma_2^2}\right)^{-1}.
\ee
Here, ${\bf e}_p = {\bf E}_p/|{\bf E}_p|$, $\tilde{\bf M}^{cv}_{\bs k} = {\bf M}^{cv}_{\bs k}/{ev_F}$ and we have defined the dimensionless parameter 
$\zeta = e|{\bf E}_p| v_F/(\hbar \omega \sqrt{\gamma_1 \gamma_2})$ - which uniquely characterizes the non-linearity in the system \cite{PhysRevB.94.195438,PhysRevB.95.155421, PhysRevB.97.045402, PhysRevB.97.205420}. See Appendix for the details of the questions. 
%
%
Equation~\eqref{n0} can be systematically expanded in powers of 
$\zeta$ with $\zeta \to 0$ denoting the equilibrium distribution (Fermi function denoted by $n^{\rm eq}_{\bs k}$) in absence of optical interactions, while 
the $\zeta^2$ terms denote the $|{\bf E}_p|^2$ correction to the modified distribution function. 
The $\zeta \to \infty$ limit is the saturation limit of maximum population inversion with $n^{(0)}_{{\bs k}}\to 0$.
The sideband population inversion corresponding to the probe frequency component $(\omega_p + \omega_s)$ \cite{Note1} can be expressed as,
\be\label{n1_k}
n^{(1)}_{\bs k} = n^{(0)}_{\bs k}\left({\bf E}_s\cdot{\bf M}^{cv}_{\bs k}\right)\left({\bf E}^*_p\cdot{\bf M}^{vc}_{\bs k}\right)\xi_{\bs k}~.
\ee
Here, we have defined, 
\bearr
& &\xi_{\bs k} = \frac{\mathcal{P}_{\bs k}}{2\hbar^2\omega^2_{\bs k}(\omega_s + i\gamma_1) +|{\bf E}_p \cdot{\bf M}^{cv}_{\bs k}|^2~\mathcal{Q}_{\bs k} }~\\
& & \mathcal{P}_{\bs k} = \frac{-(\omega_s + 2i\gamma_2)}{\left[\omega_{\bs k}-\omega_p + i\gamma_2\right]\left[\omega_{\bs k}-(\omega_p + \omega_s)-i\gamma_2\right]}~,\\
& & \mathcal{Q}_{\bs k} = \frac{2~(\omega_s + i\gamma_2)}{\left[\omega_{\bs k}-(\omega_p + \omega_s)-i\gamma_2\right]\left[\omega_{\bs k}-(\omega_p - \omega_s)+i\gamma_2\right]} \nonumber. \\
\eearr
It can easily be checked that $|n^{(1)}_{{\bs k}}|\to 0$ in both the limiting cases of vanishing intensity of the pump beam ($\zeta\to 0$) as well as in the saturation limit ($\zeta\to \infty$), as expected. 
Recall that $n_{\bs k}^{(-1)} = [n_{\bs k}^{(1)}]^*$ and the analytical expressions for the components of the inter-band coherence are presented in Appendix A.  
These $\omega_{-1}$ sideband components of the density matrix  generate a new optical sideband whose amplitude and polarization depend on the amplitude and polarization of the incident pump beam. 

\begin{figure}[t!]
\includegraphics[width = 1.01\linewidth]{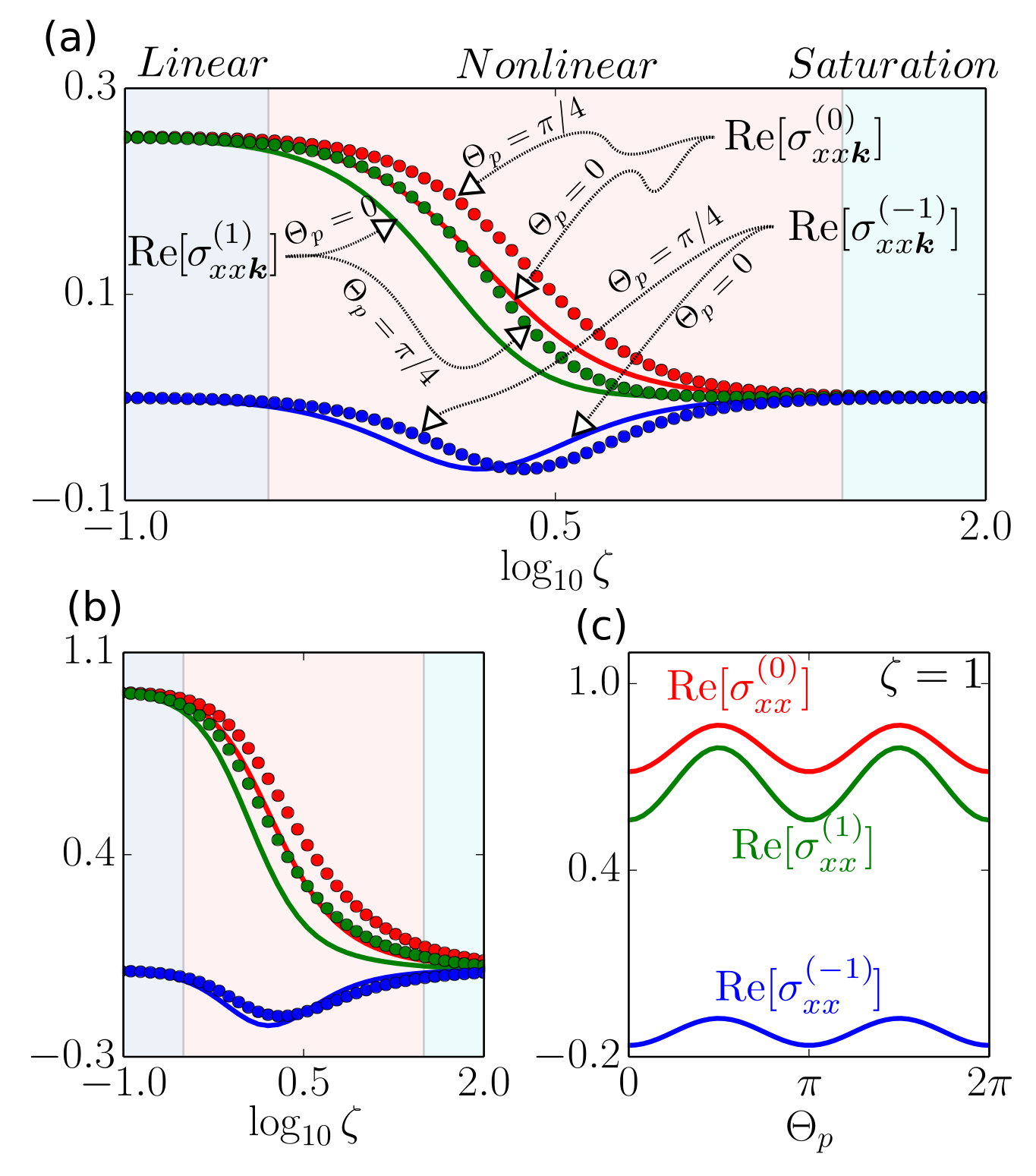}
\caption{ (a) Real part of the momentum resolved conductivity kernel corresponding to the pump ($\sigma^{(0)}_{xx\bs k}$ - in red), probe ($\sigma^{(1)}_{xx\bs k}$ - in green) and the newly generated sideband ($\sigma^{(-1)}_{xx\bs k}$ - in blue) as a function of $\log_{10}\zeta$ for $\Theta_p = 0$ (solid curve)  and $\Theta_p = \pi/4$ (dotted curve).  Here $\omega_{\bs k} = 0.8\omega_p$ and $\hat{\bs k} = (0,1)$.  
(b) The corresponding integrated optical conductivities (in units of $\sigma_0 = e^2/{4\hbar})$ as a function of $\log_{10}\zeta$. As expected, both $\sigma^{(0)}_{xx}$ and $\sigma^{(1)}_{xx}$ display linear response behaviour ($\to \sigma_0$) as $\zeta \to 0$. However, the newly generated $\sigma^{(-1)}_{xx}$ is finite only after the onset of the nonlinear response regime ($\zeta \approx 1$). 
(c) The polarization angle dependence of longitudinal conductivities for $\zeta = 1$. Other parameters are: $\omega_p = 5\times 10^{14}s^{-1}$, $\gamma_1 = 1\times 10^{12}s^{-1}$, $\gamma_2 = 5\times 10^{13}s^{-1}$ and $\mu = 0$.
\label{Fig2}}
\end{figure}%

\section{Optical conductivity} 
The optical conductivity at different frequencies can now be obtained via the calculation of the charge current response in the frequency domain: {${\bf J} = \sum_{\bs k}{\rm Trace}(\rho_{\bs k} {\bf M}_{\bs k}) = \sigma^{(0)} E_p + \sigma^{(1)} E_s + \sigma^{(-1)} E_s$, where ${\bf M}_{\bs k}/e$ is the effective velocity operator. 
Here $\sigma^{(0)}$ and $\sigma^{(\pm1)}$ are non-linear functions of $E_p$, capturing the response at frequencies $\omega_p$ and $\omega_p \pm \omega_s$, respectively.} The corresponding optical conductivity matrix can be expressed as a Brillouin zone sum of the momentum resolved 
conductivity matrix: ${\bs \sigma}^{(i)} = (2 \pi)^{-d}\int_{\rm BZ} { d \bs k}~ \sigma^{(i)}_{\bs k}$, with $i = 0, \pm 1$ and $d$ denoting the 
dimensionality of the system. 
The momentum resolved optical conductivity matrix corresponding to the pump frequency $\omega_p$ is given by 
\be\label{sigma_0}
{\bs \sigma}^{(0)}_{{\bs k}} = \frac{in^{(0)}_{{\bs k}}}{\hbar\omega_{\bs k}} 
\frac{{\bf M}^{vc}_{\bs k}{\otimes}{\bf M}^{cv}_{\bs k}}{\omega_{\bs k} - \omega'_p}. 
\ee
Here, $\omega_p' \equiv \omega_p + i \gamma_2$ and ${\otimes}$ denotes the outer product of the optical matrix element vectors. The momentum resolved optical conductivity corresponding to the probe frequency $(\omega_p +\omega_s)$ sideband is \bearr\label{sigma_1}
{\bs \sigma}^{(1)}_{{\bs k}} = {\bs \sigma}^{(0)}_{{\bs k}} \frac{1 + |{\bf E}_p\cdot{\bf M}^{cv}_{\bs k}|^2\xi_{\bf k}}{[\omega_{\bs k} - (\omega_p + \omega_s)^{'}][\omega_{\bs k} - \omega'_p]^{-1}}~.
\eearr%
Equation~\eqref{sigma_1} clearly emphasizes the gain in the optical response at the probe frequency $\omega_p+\omega_s$. As stated earlier, the response at frequencies $\omega_p$ and $(\omega_p +\omega_s)$ interfere leading to a sideband generation at the frequency $\omega_{-1} =  \omega_p - \omega_s$. The details of the calculations are presented in Appendix A and B. 

The momentum resolved optical conductivity due to the newly generated sideband is given by,  
\be\label{sigma_min_1}
{\bs \sigma}^{(-1)}_{{\bs k}} = {\bs \sigma}^{(0)}_{{\bs k}}\frac{\left({\bf M}^{vc}_{\bs k}{\otimes}{\bf M}^{cv}_{\bs k}\right)^{-1}\left({\bf M}^{vc}_{\bs k}{\otimes}{\bf M}^{vc}_{\bs k}\right) \left({\bf E}_p\cdot{\bf M}^{cv}_{\bs k}\right)^2\xi^{*}_{\bf k}}{[\omega_{\bs k} - (\omega_p - \omega_s)^{'}][\omega_{\bs k} - \omega'_p]^{-1}}~.
\ee
Equation~\eqref{sigma_min_1} highlights the optical response generated at the new sideband frequency $\omega_p - \omega_s$, and is one of the 
significant finding of this work. This new sideband response originates from the third order non-linearity in graphene. 
The dependence of the longitudinal optical conductivities on the non-linearity parameter $\zeta \propto |{\bf E}_p|$ and the pump polarization angle $\Theta_p$ 
is shown in Fig.~\ref{Fig2}. The  transverse component of the optical conductivity are presented in Fig.~\ref{transverse}. 
As expected, both $\sigma_{xx}^{(0)}$ and $\sigma_{xx}^{(1)}$ reduce to the universal optical conductivity of graphene, 
$\sigma_0 = e^2/(4 \hbar)$, in the linear response regime of $\zeta \to 0$. However, the new sideband contribution $\sigma_{xx}^{(-1)}$ is finite only in the non-linear regime of $\zeta \approx 1$, and vanishes in the linear response as well as in the saturation regime ($\zeta \gg  1$). 

{\begin{table}[t] 
\caption{The reflection coefficients in two dimensional materials in terms of  optical conductivities~\cite{Yoshino:13, PhysRevB.97.205420}. Here we have defined, ${\tilde\sigma}_{ij} = {\sigma}_{ij}/\sigma_0$, $\sigma_0 = e^2/(4 \hbar)$, and $\tilde{Z} = 2/(\pi\alpha_F \tilde{\sigma}_d)$ where $\alpha_F \approx 1/137$ is the fine structure constant, and $\tilde{\sigma}_d = \left(2/{\pi\alpha_F} + {\tilde \sigma}_{xx}\right)\left(2/{\pi\alpha_F} + {\tilde \sigma}_{yy}\right)
-{\tilde \sigma}_{xy}^2$ \cite{Note2}.  \label{T1}}
\vspace{0.2cm}
\begin{tabular}{ c c c c c} \hline \hline 
Coefficient &  Exact expression &  $ {\cal O} (\alpha_F)$ \\ \hline
$r_{ss}$ & $\tilde{Z}\left(\tilde{Z}\tilde{\sigma}_d + { \tilde \sigma}_{yy}\right)-1$ &~~ $-{\tilde\sigma}_{xx}/(\tilde{Z}\tilde{\sigma}_d)$ &  \\ 
$r_{pp}$ & $1-\tilde{Z}\left(\tilde{Z}\tilde{\sigma}_d + \tilde{\sigma}_{xx}\right)$ &~~ ~~${\tilde\sigma}_{yy}/(\tilde{Z}\tilde{\sigma}_d)$&  \\
$r_{sp}$ &$-\tilde{Z}{\tilde \sigma}_{xy}$ & ~~~$-{\tilde\sigma}_{xy}/(\tilde{Z}\tilde{\sigma}_d)$&   \\ 
$\chi_{s\rm Kerr}$ & ~$-r_{ps}/r_{ss}$ &~~ ${\sigma}_{xy}/{\sigma}_{xx}$ &  \\ 
$\chi_{p\rm Kerr}$ & ~~~$r_{sp}/r_{pp}$ &~~ $-{\sigma}_{xy}/{\sigma}_{yy}$ &  
\\ 
 \hline
\end{tabular} 
\end{table}
%


\section{Experimental implications of the new sideband} 
The generated sideband would leave its signature in a range of optical and photo-conductivity measurements \cite{Prechtel2012}.  Here, we focus on its impact in optical reflectivity, In particular, we explore 
the pump power and polarization angle dependence of the reflectivity (amplitude and phase in $r = \sqrt{\cal R}e^{i \Phi}$), and the Kerr angle $\Theta_{\rm Kerr}$ \cite{Yoshino:13, PhysRevB.97.205420}. For graphene, though small, the reflectivity is routinely measured \cite{Nair2008,Mak,PhysRevB.95.155421,PhysRevB.97.205420}, while the phase of the reflection coefficient can be measured using a generic interference setup. Thus ${\cal R}$, $\Phi$ and $\Theta_{\rm Kerr}$ can be probed as a function of the probe laser power and polarization angle (see Fig.~\ref{Fig1}). The dependence of the $s$ and $p$ components of the reflection coefficients on the respective optical conductivities in graphene are tabulated in Table \ref{T1} \footnote{The simplified expressions in the last column are obtained using $\alpha_F\ll 1$, along with $\sigma^2_{xy} \ll \sigma_{0}$ -- which works in the case of graphene.}.

\begin{figure}[t!]
\includegraphics[width = 1.01\linewidth]{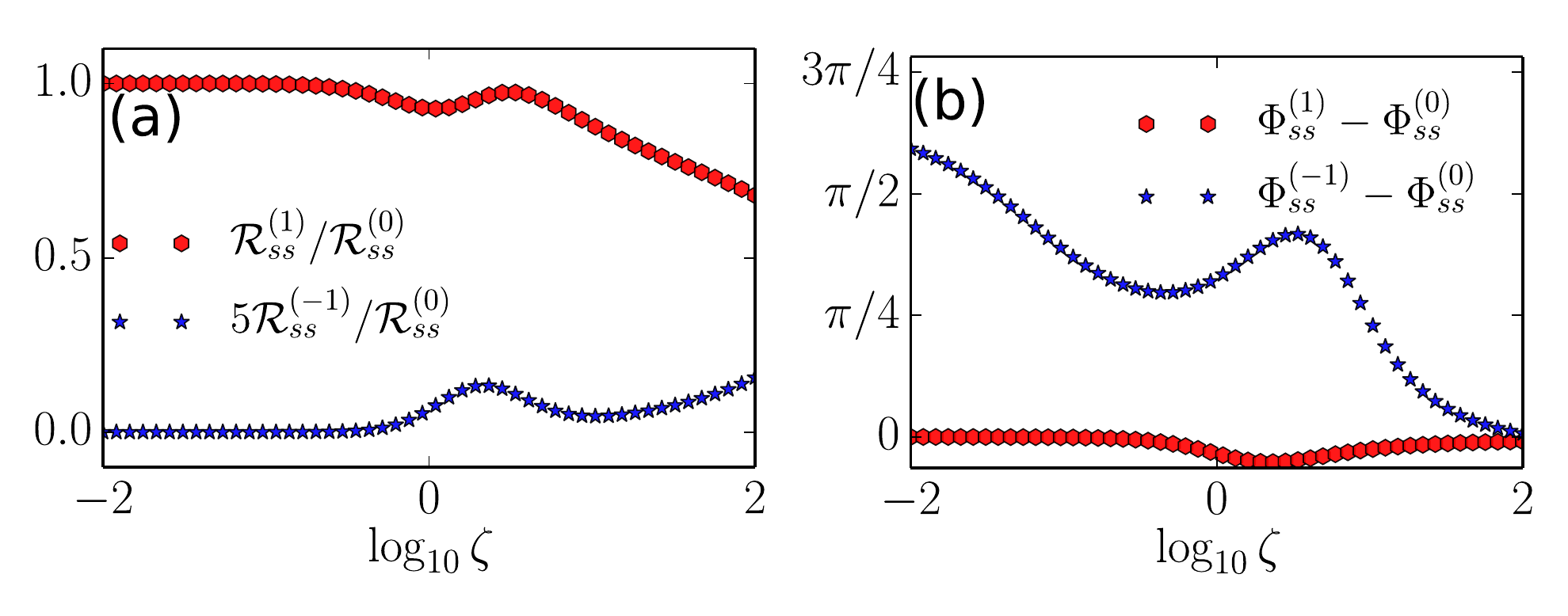}
\caption{Variation of (a) reflection probability (b) and its phase as a function of the pump field strength ($\log_{10}\zeta$), for the probe, and $\omega_{-1}$ frequencies. 
The nonlinear $\omega_{-1}$ sideband can also be probed via the interference experiments sensitive to the phase of the reflection coefficient.  
We have chosen $\Theta_p = \pi/4$ for both panels, and other parameters are identical to those of Fig.~\ref{Fig2}.}
\label{reflectance}
\end{figure}

To compare the reflection amplitude and phase of the sideband 
\footnote{ Reflectivity and polarization rotation for the newly generated sideband are defined in reference to the amplitude of backward propagating field $r^{(-1)}$, and its polarization relative to the polarization of the incident pump beam.}  
with that of the pump beam, we define the following: 
\be\label{Refl_phase}
\frac{\mathcal{R}^{(\lambda)}_{ss}}{\mathcal{R}^{(0)}_{ss}} = \frac{|r^{(\lambda)}_{ss}|^2}{|r^{(0)}_{ss}|^2}~,~~~~\tan\Phi^{(\lambda)} = \frac{{\rm Im}[r^{(\lambda)}_{ss}]}{{\rm Re}[r^{(\lambda)}_{ss}]}~,
\ee
where $\lambda = \pm 1$ for reflectance measured at the sideband frequencies $\omega_p \pm \omega_s$. 
The dependence of the ratios of the reflectance and $\Phi$ defined in Eq.~\ref{Refl_phase}, is shown in Fig.~\ref{reflectance} as a function of the pump beam intensity ($\propto \zeta^2$). Clearly the $\omega_{-1}$ sideband response at manifests only in the non-linear regime of $\zeta \approx 1$. In the optical regime (say $\omega_p = 5\times10^{14}{\rm s^{-1}}$),  the estimated damping constants in graphene \cite{Zhang1} are $\gamma_1 \approx 10^{12}{\rm s^{-1}}, \gamma_2 \approx 5\times10^{13}{\rm s^{-1}}$. Using these values, the $\zeta \approx 1$ condition in graphene corresponds to a CW laser intensity $\approx 10^5~{\rm Wcm^{-2}}$, which is reasonable \cite{Prechtel2012}.
Furthermore, at reasonable CW powers we also observe non-linear phase shifts in excess of $\pi/2$ for the new sideband. Such large non-linear phase shifts is of great interest in a range of switching applications in THz and optical domains. The polarization angle $\Theta_p$ dependence of the reflection probability and its phase is shown in Fig.~\ref{Kerr}(a)-(b). 

Non-linear optical response in graphene also generates a finite $\sigma_{xy}$, which in turn leads to 
Kerr rotation (polarization rotation of the reflected beam)\cite{Yoshino:13, PhysRevB.97.205420}.
The Kerr rotation angle for $s-$ and $p-$ polarized incident pump beam is given by \cite{Yoshino:13, PhysRevB.97.205420},
\be
\Theta^{(\lambda)}_{s/p{\rm Kerr}} = \frac{1}{2}\tan^{-1}\left(\frac{2{{\rm Re}[\chi^{(\lambda)}_{s/p{\rm Kerr}}]}}{1 - |\chi^{(\lambda)}_{s/p{\rm Kerr}}|^2}\right)~,
\ee%
where $\chi_{s/p{\rm Kerr}}$ can be expressed in terms of the reflection coefficients (see Table~\ref{T1}). 
The variation of the Kerr angle for the $s$ and $p$ components for pump, probe and the new sideband beam as a function of $\Theta_p$ is shown in Fig.~\ref{Kerr} (c)-(d). The polarization rotation of the $\omega_{-1}$ sideband seems to be significantly large and different from that corresponding to the pump  and probe frequencies. 
%
\begin{figure}[t]
\includegraphics[width = 1.01\linewidth]{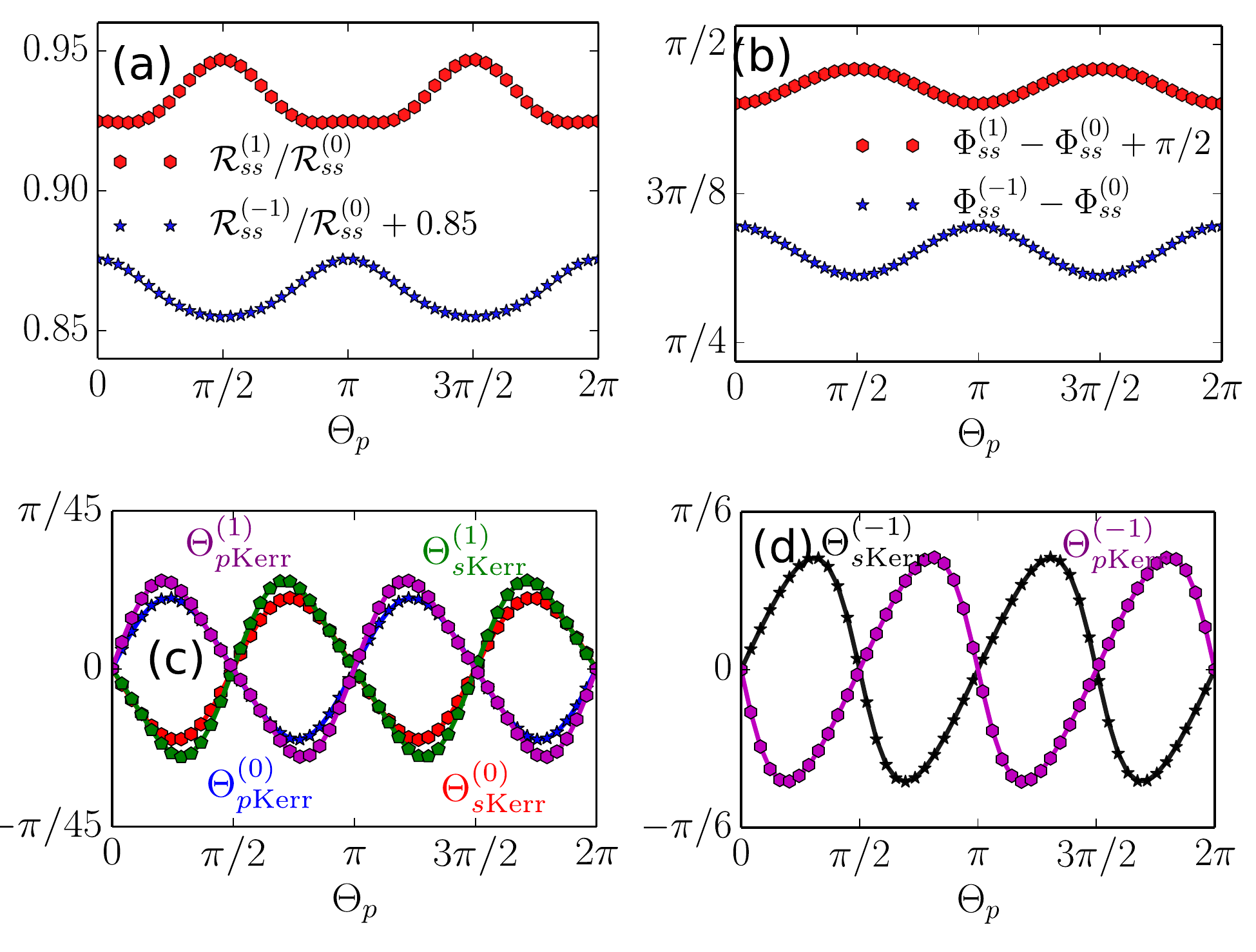}
\caption{Dependence of (a) the probability and (b) the phase of the reflection coefficient for the probe and $\omega_{-1}$ sideband as a function of $\Theta_p$.
The polarization rotation of the reflected beam (Kerr angle) for $s-$ and $p-$ polarized beams as a function of $\Theta_p$ are shown in c) for the pump and probe frequencies and in d) for the $\omega_{-1}$ frequency. Here $\zeta = 1$ and 
other parameters are identical to those of Fig.~\ref{Fig2}.}
\label{Kerr}
\end{figure}
%
\section{Summary}
In summary, we predict generation of a new modulated optical sideband in graphene in presence of a CW frequency {shifted} pump-probe setup. Physically, the  `slushing' of the inter-band coherence due to interference of the pump and the probe results in the generated sideband that carries unique signature of the third order non-linear response in graphene. Experimentally, this manifests in the polarization, reflectivity, and in the phase of the reflection coefficient (see Fig.~\ref{reflectance}) at the sideband frequencies. 
In particular, the peak of the sideband gain occurs at a thereshold, characterized by a single parameter $\zeta$ set by system decay rates and the pump power. A careful characterization of generated sideband gain can thereby provide a direct method of characterizing non-linear response of two-band systems with CW fields, in contrast to traditional, technologically involved time domain measurements. It also suggests a range of applications that include switching of frequency sidebands using non-linear phase shifts and generation of frequency combs.}

\appendix 

\section{Steady state density matrix in presence of pump and probe fields}

\begin{figure*}[t!]
\includegraphics[width = 0.65\linewidth]{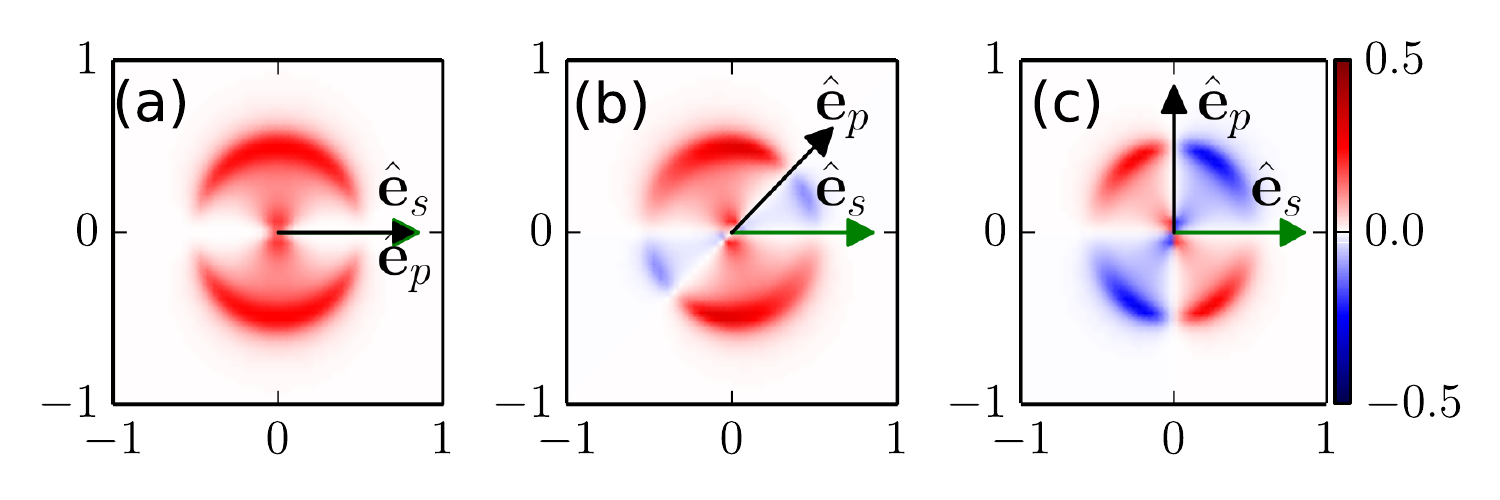}
\caption{Real part of population inversion corresponding to probe frequency ($\Re [n_{\bf k}^{(1)}]$) for, (a) $\hat{\bf e}_p = \hat{\bs x}$, $\hat{\bf e}_s = \hat{\bs x}$, (b) $\hat{\bf e}_p = (\hat{\bs x} + \hat{\bs y})/\sqrt{2}$, $\hat{\bf e}_s = \hat{\bs x}$ and (c) $\hat{\bf e}_p = \hat{\bs y}$, $\hat{\bf e}_s = \hat{\bs x}$. Other parameters are: $\omega_p = 5\times 10^{14}s^{-1}$, $\gamma_1 = 1\times 10^{12}s^{-1}$, $\gamma_2 = 5\times 10^{13}s^{-1}$ and the chemical potential $\mu = 0$.
\label{pop_inv}}
\end{figure*}%

In this section we obtain analytical results for the steady state density matrix as a solution of the optical Bloch equations (OBEs) in the presence of a pump as well as a probe field. 
The OBEs, including phenomenological damping terms are given by~\cite{PhysRevB.94.195438,PhysRevB.95.155421, PhysRevB.97.045402, PhysRevB.97.205420},
\bearr\label{OBE_App}
\partial_t n_{\bs k} &=& 4~{\rm Im}\left[{\Omega}^{cv}_{\bs k}p_{\bs k}\right] - \gamma_1\left(n_{\bs k}-n^{\rm eq}_{\bs k}\right),\\
\partial_t p_{\bs k} &=& i\omega_{\bs k} p_{\bs k} - i{\Omega}^{vc}_{\bs k}n_{\bs k} - \gamma_2 p_{\bs k}\label{OBE_p}~. 
\eearr
The ansatz for the solution of population inversion, $n_{\bs k}$ and the inter-band coherence $p_{\bs k}$ is motivated by the fact that the relatively weak probe field has only perturbative impact on the steady state population inversion achieved under the action of the pump field alone. Following Boyd\cite{Boyd}, we can express $n_{\bs k}$ and $p_{\bs k}$ as,
\bearr\label{Ans_n1}
n_{\bs k} &=& n^{(0)}_{\bs k} + n^{(1)}_{\bs k} e^{-i\omega_s t} + n^{(-1)}_{\bs k} e^{i\omega_s t}~, \\
p_{\bs k} &=& \left(p^{(0)}_{\bs k} + p^{(1)}_{\bs k}e^{-i\omega_s t} + p^{(-1)}_{\bs k}e^{i\omega_s t}\right) e^{i\omega_p t}~ \nn,
\eearr
where $n^{(0)}_{\bs k}, n^{(1)}_{\bs k}$ and $n^{(-1)}_{\bs k}$ are time independent in the steady state. Here we assume that $\omega_s\ll\omega_p$ and we ignore the second order terms like $n^{(1)}p^{(1)}_{\bs k}, n^{(1)}p^{(-1)}_{\bs k}$ and so on. Since the total population inversion $n_{\bs k}$ has to be a real physical quantity, we have $({n^{(1)}_{\bs k}})^* = n^{(-1)}_{\bs k}$.

The time derivative of the population inversion and the inter-band polarization are given by,
\bearr\label{dot_n}
\dot{n}_{\bs k} &=& -i\omega_s~n^{(1)}_{\bs k}e^{-i\omega_s t} + i\omega_s~n^{(-1)}_{\bs k}e^{i\omega_s t}~, \\\nn
\dot{p}_{\bs k} &=& i\omega_p~p^{(0)}_{\bs k}e^{i\omega_p t} + i(\omega_p-\omega_s)~p^{(1)}_{\bs k}e^{i(\omega_p -\omega_s)t}\label{dot_p}\\
&+& i(\omega_p+\omega_s)~p^{(-1)}_{\bs k}e^{i(\omega_p +\omega_s)t} ~.
\eearr
Using the expression for $\Omega^{vc}_{\bs k}$ and the full form of the applied pump and probe electric field, 
a straightforward calculation yields,
\bearr\label{Omega_n}\nn
-i\Omega^{vc}_{\bs k}{n}_{\bs k} &=& 
 \Bigg[\frac{{\bf E}_p\cdot{\bf M}^{vc}_{\bs k} }{{2\hbar\omega_{\bs k}}}n^{(0)}_{\bs k}e^{-i\omega_p t} + \frac{{\bf E}^*_p\cdot{\bf M}^{vc}_{\bs k} }{{2\hbar\omega_{\bs k}}}n^{(0)}_{\bs k}e^{i\omega_p t}\\\nn
&+& \Big(\frac{{\bf E}_p\cdot{\bf M}^{vc}_{\bs k} }{{2\hbar\omega_{\bs k}}}n^{(1)}_{\bs k}
+ \frac{{\bf E}_s\cdot{\bf M}^{vc}_{\bs k} }{{2\hbar\omega_{\bs k}}}n^{(0)}_{\bs k}\Big)e^{-i(\omega_p + \omega_s)t}\\\nn
&+& \Big(\frac{{\bf E}^*_s\cdot{\bf M}^{vc}_{\bs k} }{{2\hbar\omega_{\bs k}}}n^{(0)}_{\bs k}
+ \frac{{\bf E}^*_p\cdot{\bf M}^{vc}_{\bs k} }{{2\hbar\omega_{\bs k}}}n^{(-1)}_{\bs k}\Big)e^{i(\omega_p + \omega_s)t}\\\nn
&+& \frac{{\bf E}_p\cdot{\bf M}^{vc}_{\bs k} }{{2\hbar\omega_{\bs k}}}n^{(-1)}_{\bs k}e^{-i(\omega_p - \omega_s)t}\\
&+& \frac{{\bf E}^*_p\cdot{\bf M}^{vc}_{\bs k} }{{2\hbar\omega_{\bs k}}}n^{(1)}_{\bs k}e^{i(\omega_p - \omega_s)t}
\Bigg].
\eearr
Ignoring the counter rotating terms in Eq.~\eqref{Omega_n} and using Eq.~\eqref{dot_p}, we obtain
\be\label{p0_k}
p^{(0)}_{\bs k} = \frac{i}{2\hbar\omega_{\bs k}}\left[\frac{n^{(0)}_{\bs k}\left({\bf E}^*_p\cdot{\bf M}^{vc}_{\bs k}\right)}{\omega_{\bs k} - \omega_p + i\gamma_2}\right],
\ee
\be\label{p1_k}
p^{(1)}_{\bs k} = \frac{i}{2\hbar\omega_{\bs k}}\left[\frac{n^{(1)}_{\bs k}\left({\bf E}^*_p\cdot{\bf M}^{vc}_{\bs k}\right)}{\omega_{\bs k} - (\omega_p-\omega_s) + i\gamma_2}\right],
\ee
and,
\be\label{pmin1_k}
p^{(-1)}_{\bs k} = \frac{i}{2\hbar\omega_{\bs k}}\left[\frac{\left({\bf E}^*_p\cdot{\bf M}^{vc}_{\bs k}\right)n^{(-1)}_{\bs k} + \left({\bf E}^*_s\cdot{\bf M}^{vc}_{\bs k}\right)n^{(0)}_{\bs k}}{\omega_{\bs k} - (\omega_p + \omega_s) + i\gamma_2}\right].
\ee
Note that we have done the calculations keeping the counter-rotating terms as well, and explicitly checked that the results are qualitatively in very good agreement to the ones reproduced here. 

Similar to Eq.~\eqref{Omega_n}, we have,
\bearr\label{Omega_p}\nn
\Omega^{cv}_{\bs k}p_{\bs k} &=& -i\frac{{\bf E}_p\cdot{\bf M}^{cv}_{\bs k}}{2\hbar\omega_{\bs k}}p^{(0)}_{\bs k}-i\frac{{\bf E}_p\cdot{\bf M}^{cv}_{\bs k}}{2\hbar\omega_{\bs k}}p^{(-1)}_{\bs k}e^{i\omega_s t}\\
&-& i\left(\frac{{\bf E}_p\cdot{\bf M}^{cv}_{\bs k}}{2\hbar\omega_{\bs k}}p^{(1)}_{\bs k} + \frac{{\bf E}_s\cdot{\bf M}^{cv}_{\bs k}}{2\hbar\omega_{\bs k}}p^{(0)}_{\bs k}\right)e^{-i\omega_s t}~. \nonumber \\
\eearr
Substituting Eq.~\eqref{Omega_p} in Eq.~\eqref{OBE_App} we obtain,
\be
\gamma_1\left(n^{(0)}_{\bs k} - n^{\rm eq}_{\bs k}\right) = 
-\frac{{\bf E}_p\cdot{\bf M}^{cv}_{\bs k}}{\hbar\omega_{\bs k}}p^{(0)}_{\bs k} - \frac{{\bf E}^*_p\cdot{\bf M}^{vc}_{\bs k}}{\hbar\omega_{\bs k}}{p^{(0)}_{\bs k}}^*.
\ee
Combining this with Eq.~\eqref{p0_k}, we obtain 
\be\label{n0_k}
\frac{n^{(0)}_{\bs k}}{n^{\rm eq}_{\bs k}} = \left[1 + \frac{\gamma_2}{\gamma_1}\frac{1}{\hbar^2\omega_{\bs k}^2}\frac{|{\bf E}_p \cdot{\bf M}^{cv}_{\bs k}|^2}{(\omega_{\bs k}-\omega_p)^2 + \gamma_2^2} \right]^{-1}.
\ee%
%
%
%
%
%
%
Similarly, the population inversion corresponding to the probe frequency can be obtained to be
\bearr\nn
(\gamma_1 - i\omega_s)n^{(1)}_{\bs k} &=& 
-\frac{{\bf E}_p\cdot{\bf M}^{cv}_{\bs k}}{\hbar\omega_{\bs k}}p^{(1)}_{\bs k} - \frac{{\bf E}_s\cdot{\bf M}^{cv}_{\bs k}}{\hbar\omega_{\bs k}}p^{(0)}_{\bs k} \\
&-&\frac{{\bf E}^*_p\cdot{\bf M}^{vc}_{\bs k}}{\hbar\omega_{\bs k}}{p^{(-1)}_{\bs k}}^*.
\eearr
Combining this with Eq.~\eqref{p0_k}, Eq.~\eqref{p1_k} and Eq.~\eqref{pmin1_k}, leads to 
\be
n^{(1)}_{\bs k} = n^{(0)}_{\bs k}\left({\bf E}_s\cdot{\bf M}^{cv}_{\bs k}\right)\left({\bf E}^*_p\cdot{\bf M}^{vc}_{\bs k}\right)\xi_{\bs k}~.
\ee
Here we have defined 
\be
\xi_{\bs k} = \frac{\mathcal{P}_{\bs k}}{2\hbar^2\omega^2_{\bs k}(\omega_s + i\gamma_1) +|{\bf E}_p \cdot{\bf M}^{cv}_{\bs k}|^2~\mathcal{Q}_{\bs k} }~,
\ee
\be
\mathcal{P}_{\bs k} = \left(\frac{1}{\omega_{\bs k}-\omega_p + i\gamma_2}-\frac{1}{\omega_{\bs k}-(\omega_p + \omega_s) - i\gamma_2}\right),
\ee
and 
\be
\mathcal{Q}_{\bs k} = \left(\frac{1}{\omega_{\bs k}-(\omega_p + \omega_s) - i\gamma_2}-\frac{1}{\omega_{\bs k}-(\omega_p - \omega_s) + i\gamma_2}\right).
\ee
The real part of the obtained $n^{(1)}_{\bs k}$ is shown in Fig.~\ref{pop_inv}, for different orientation of the polarization angles of the pump beam. 

\section{Pump, probe and the sideband current density}

In this section we calculate the inter-band current density corresponding to the pump, probe and the newly generated sideband frequencies.
In presence of a frequency modulated CW light beam, a steady state situation is achieved where a quasi stationary population inversion is obtained as shown explicitly in the previous section. 
During this period, a non-vanishing steady state inter-band current is maintained because of the finite inter-band coherence or polarization.
The momentum resolved current density, 
at any time $t$ can be expressed in terms of microscopic polarization $p_{\bs k}$ and the optical matrix element ${\bf M}^{cv}_{\bs k}$
as,
\be \label{def_J0}
{\bf J}_{\bs k}(t) =  -2 {\rm Re}[p_{\bs k}(t) {\bf M}^{cv}_{\bs k}]~.
\ee
The total current is given as the sum of all the momentum modes over the Brillouin zone (BZ),
\be\label{int_J0}
{\bf J}(t) = \frac{g_sg_v}{4\pi^2}\int {\bf J}_{\bs k}(t)d{\bs k},
\ee
where $g_s$ and $g_v$ represents the spin and valley degeneracy respectively. 
While using a tight-binding model, this summation is restricted to the first Brillouin zone. On using an effective low energy model, the integral limit is fixed to some cut-off value where the integral kernel is almost zero. 
In the presence of the probe sideband at $\omega_p + \omega_s$ frequency, we need to make a similar {\it ansatz} for the total current as we did for the components of the density matrix. Thus the total current can be expressed as,
\be
 {\bf J}_{\bs k}(t) = \frac{1}{2}\left[{\bs J}_{\bs k} e^{-i \omega_p t} + c.c\right],
\ee
 where $
{\bs J}_{\bs k} = {\bs J}^{(0)}_{\bs k} + {\bs J}^{(1)}_{\bs k}e^{-i\omega_s t} + {\bs J}^{(-1)}_{\bs k}e^{i\omega_s t}.
$
Note that the different time dependence of these currents will lead to the generation of electromagnetic fields at different optical frequencies: $\omega_p$ and $\omega_p \pm \omega_s$. 
Once again, matching the coefficient of $e^0, e^{i\omega_s t}$ and $e^{-i\omega_s t}$ terms, we obtain ${\bs J}^{(0)}_{\bs k} = -2~{p^{(0)}_{\bs k}}^*{\bf M}^{vc}_{\bs k}, {\bs J}^{(1)}_{\bs k} = -2~{p^{(-1)}_{\bs k}}^*{\bf M}^{vc}_{\bs k}$ and ${\bs J}^{(0)}_{\bs k} = -2~{p^{(1)}_{\bs k}}^*{\bf M}^{vc}_{\bs k}$. This implies that,
\be
{\bs J}^{(0)}_{\bs k} = \frac{in^{(0)}_{\bs k}}{\hbar\omega_{\bs k}}\frac{{\left({\bf E}_p\cdot{\bf M}^{cv}_{\bs k}\right){\bf M}^{vc}_{\bs k}}}{\omega_{\bs k} - \omega_p - i\gamma_2}~,
\ee
\be
{\bs J}^{(1)}_{\bs k} = \frac{{\left({\bf E}_s\cdot{\bf M}^{cv}_{\bs k}\right){\bf M}^{vc}_{\bs k}}}{\hbar\omega_{\bs k}}\left[\frac{in^{(0)}_{\bs k}\left(1 + \xi_{\bs k}|{\bf E}_p\cdot{\bf M}^{cv}_{\bs k}|^2\right)}{\omega_{\bs k} - (\omega_p+\omega_s) - i\gamma_2}\right]~,
\ee
and,
\be
{\bs J}^{(-1)}_{\bs k} = \frac{in^{(0)}_{\bs k}}{\hbar\omega_{\bs k}}\left[\frac{\left({\bf E}^*_s\cdot{\bf M}^{vc}_{\bs k}\right){\bf M}^{vc}_{\bs k}{\left({\bf E}_p\cdot{\bf M}^{cv}_{\bs k}\right)^2}\xi^*_{\bs k}}{\omega_{\bs k} - (\omega_p-\omega_s) - i\gamma_2}\right]~.
\ee
We consider the optical electric field associated with the pump and probe beams to be real i.e., ${\bf E}_p = |{\bf E}_p|\left(\cos\Theta_p, \sin\Theta_p\right)$, and ${\bf E}_s = |{\bf E}_s|\left(\cos\Theta_s, \sin\Theta_s\right)$, where $|{\bf E}_p|$ and $|{\bf E}_s|$ are the magnitudes and, $\Theta_p$ and $\Theta_s$ are the polarization direction of the respective beams. 

\begin{figure*}[ht!]
\includegraphics[width = 0.65\linewidth]{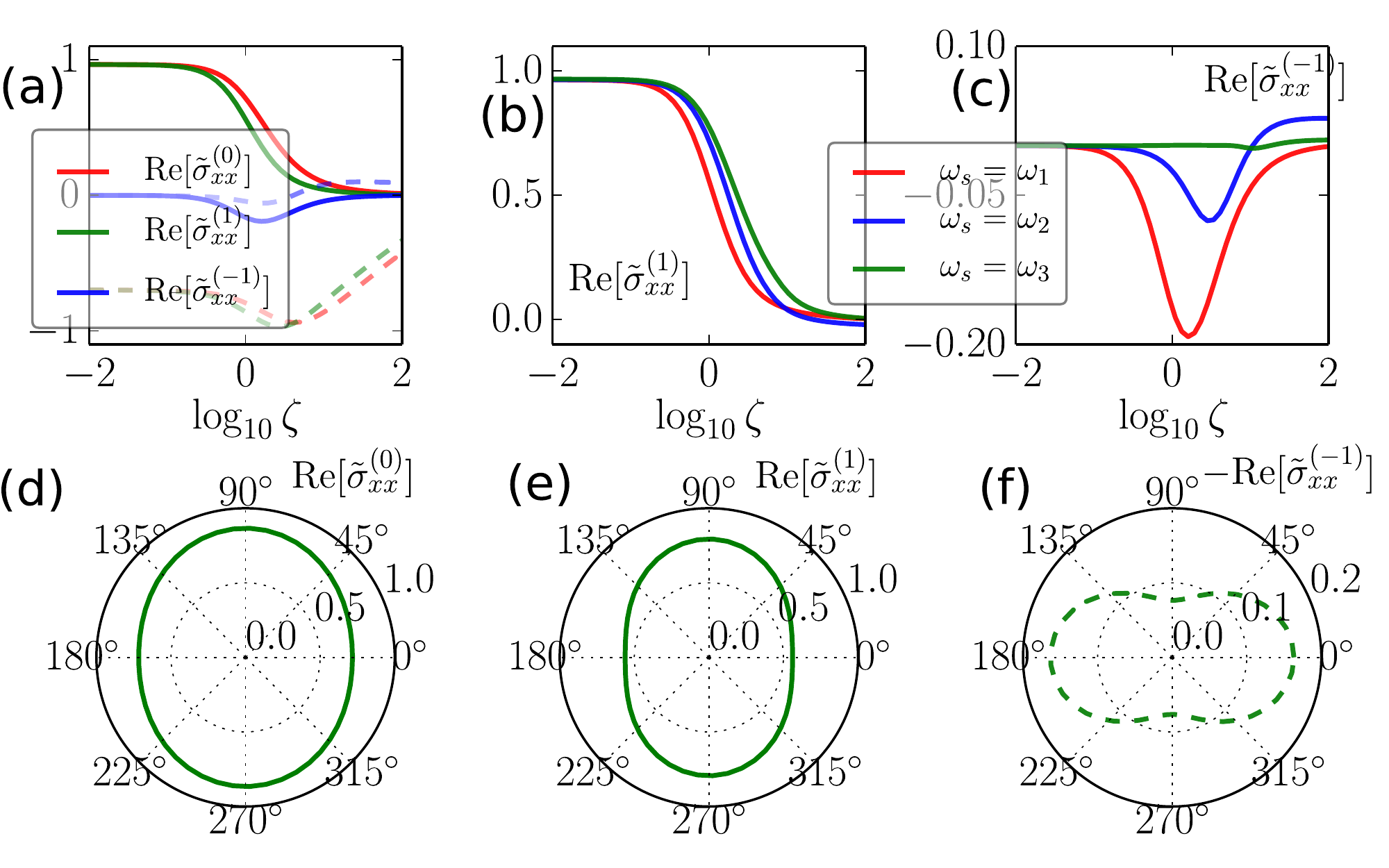}
\caption{Longitudinal optical conductivities (scaled with $\sigma_0 = e^2/(4\hbar)$) as a function of electric field dependend dimensionless parameter $\zeta$ and polarization. 
(a) Real (shown in solid) and imaginary part (shown in dashed) corresponding to the pump ($\sigma^{(0)}_{xx}$), probe ($\sigma^{(1)}_{xx}$) and the newly generated sideband
($\sigma^{(-1)}_{xx}$) as a function of $\log_{10}\zeta$ for $\Theta_p = 0$. (b) and (c) demonstrate the response of $\sigma^{(1)}_{xx}$ and $\sigma^{(-1)}_{xx}$ for three 
different values of the sideband frequencies, $10^{11}{\rm s^{-1}}$, $5\times10^{12}{\rm s^{-1}}$ and $5\times10^{13}{\rm s^{-1}}$, showing that the former increases while 
the latter decreases with increase in $\omega_s$. In (d), (e), and (f) we have shown the dependence of $\sigma^{(0)}_{xx}$, $\sigma^{(1)}_{xx}$ and $\sigma^{(-1)}_{xx}$ w.r.t. $\Theta_p$ for $\zeta = 1$. Please note that we have shown ${\rm Re}[\sigma^{(-1)}_{xx}]$ in dashed in Fig. \ref{conductivity}(f), because of -ve sign, as can also be seen from Fig. \ref{conductivity}(c). 
Other parameters are same as those of Fig. \ref{pop_inv}.
\label{conductivity}}
\end{figure*}
%
%

 The Hamiltonian of Graphene in the Fourier space is given as~\cite{katsnelson2012graphene}
\be \label{H_gr}
H = \begin{pmatrix}
     0 && tf({\bs k})\\
     tf^*({\bs k})&& 0 \\
    \end{pmatrix}~,
\ee
where the hopping parameter $t$ is roughly equal to $2.7$ eV, 
$f({\bs k}) = e^{i{\bs k}\cdot{\boldsymbol{\delta}_1}} + e^{i{\bs k}\cdot{\boldsymbol{\delta}_2}} + e^{i{\bs k}\cdot{\boldsymbol{\delta}_3}}$,~
$\boldsymbol{\delta}_1 = \frac{a}{2}\left(1, \sqrt{3}\right)$, $\boldsymbol{\delta}_2 = \frac{a}{2}\left(1, -\sqrt{3}\right)$,  
$\boldsymbol{\delta}_3 = a\left(-1, 0\right)$, $f^*({\bs k})$ is the complex conjugate of $f({\bs k})$ and ${\bs k} = (k_x, k_y)$.
Thus we have,
\bearr\nn
f({\bs k}) = \cos\left[k_xa\right] + 2\cos\left[{k_xa}/{2}\right]\cos\left[{\sqrt{3}k_ya}/{2}\right]\\
           -i\left(\sin\left[k_xa\right] - 2\sin\left[{k_xa}/{2}\right]\cos\left[{\sqrt{3}k_ya}/{2}\right]\right).
\eearr
The energy eigenvalues of Eq.~\eqref{H_gr} are given by $\varepsilon^{\pm}_{\bs k} = \pm E_{\bs k}$, where $E_{\bs k} = t|f({\bs k})|$. The bandstructure of graphene apparently has 6 Dirac points. However out of these six points, only a pair of Dirac points are not equivalent and are usually referred as ${\bf K}$ and ${\bf K}'$ points. Generally these are chosen to be located at 
\be\label{KK'}
{\bf K} = \left(\frac{2\pi}{3a}, -\frac{2\pi}{3\sqrt{3}a}\right), ~~{\rm and}~~{\bf K}' = \left(\frac{2\pi}{3a}, \frac{2\pi}{3\sqrt{3}a}\right)~.
\ee
The low energy dispersion of graphene close to either of these two Dirac points is given  by,
\be\label{Gra_low}
H \approx \begin{pmatrix}
     0 && \hbar v_F\left(k_x - ik_y\right)\\
     \hbar v_F\left(k_x + ik_y\right) && 0 \\
    \end{pmatrix}.
\ee
Here, $k$ is measured from either of the Dirac points. For this effective low energy Hamiltonian, the energy eigenvalue is given by $E_{\bs k} = \hbar v_F |{\bs k}|$,  where the Fermi velocity is known to be $v_F = 10^6{\rm ms^{-1}}$ and $|{\bs k}| = \sqrt{k_x^2 + k_y^2}$~. 

The optical matrix element responsible for the inter-band transition in the generic two band model is \cite{PhysRevB.95.155421},
\bearr\label{Mvc}
{\bf M}_{\bs k}^{vc} &=& -\frac{e}{\hbar g_{\bs k} h_{\bs k}}\Big(- h_{\bs k}^2\nabla_{\bs k} h_{3 {\bs k}}  + (h_{1 {\bs k}} h_{3 {\bs k}} - i h_{2 {\bs k}} g_{{\bs k}})  \nabla_{\bs k} h_{1 {\bs k}} \nn \\
 &+&  (h_{2 {\bs k}} h_{3 {\bs k}} + i h_{1 {\bs k}} g_{{\bs k}})  \nabla_{\bs k} h_{2 {\bs k}}  ~\Big).
\eearr%
\begin{figure*}[ht!]
\includegraphics[width = 0.65\linewidth]{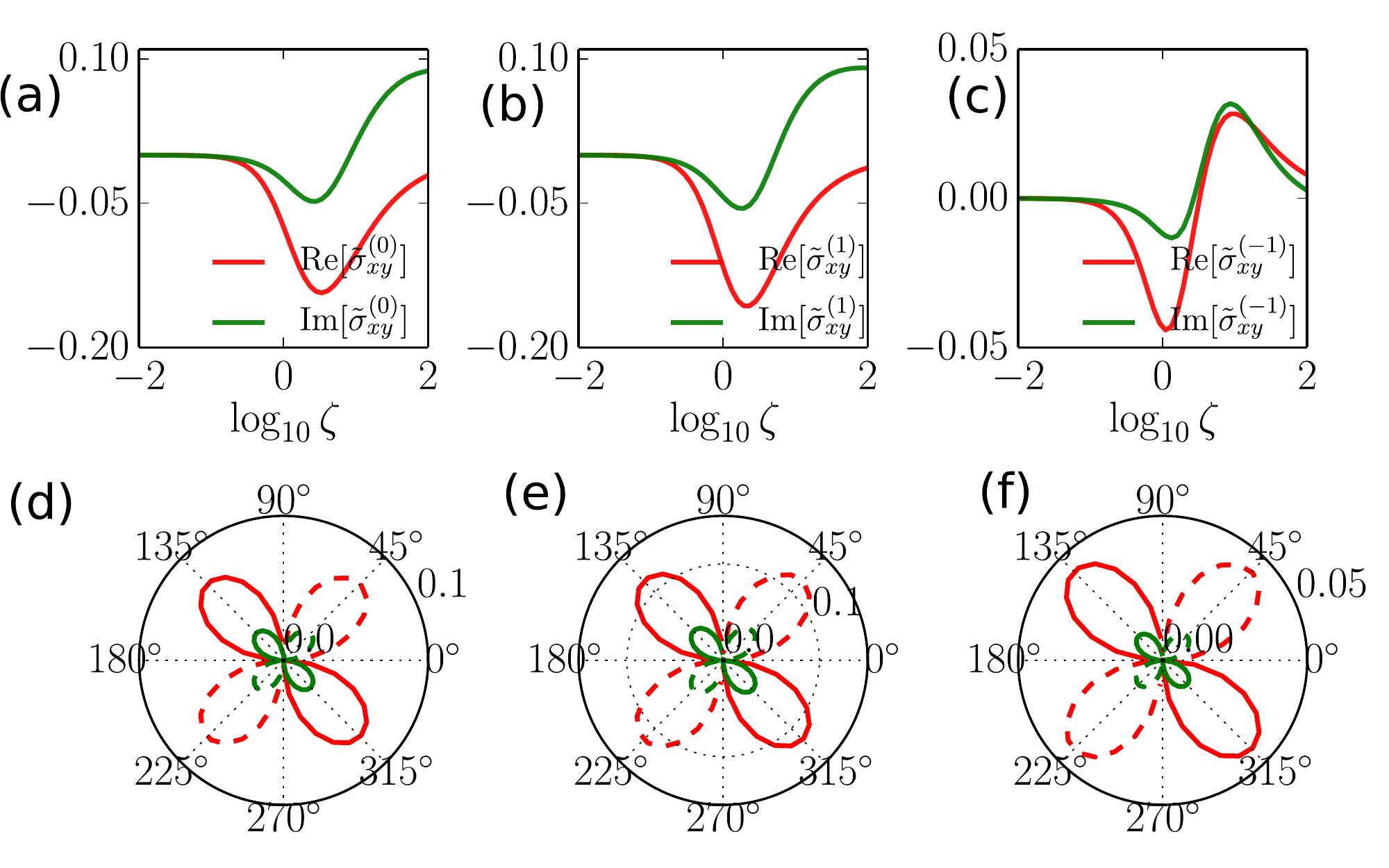}
\caption{Real and imaginary parts of transverse optical conductivities (scaled with $\sigma_0 = e^2/(4\hbar)$) as a function of electric field dependent dimensionless parameter $\zeta$ and polarization. We have shown
(a)  {\rm Re}[$\sigma^{(0)}_{xy}$] and {\rm Im}[$\sigma^{(0)}_{xy}$], (b)  {\rm Re}[$\sigma^{(1)}_{xy}$] and {\rm Im}[$\sigma^{(1)}_{xy}$] and (c) {\rm Re}[$\sigma^{(-1)}_{xy}$] and {\rm Im}[$\sigma^{(-1)}_{xy}$] as a function of $\log_{10}\zeta$ for $\Theta_p = \pi/4$ whereas in (d), (e), and (f) we have shown the same set of variables w.r.t. $\Theta_p$ for $\zeta = 1$. Please note that the dashed curves represent the negative values of the variables.
Other parameters are same as those of Fig. \ref{pop_inv}.
\label{transverse}}
\end{figure*}%
Specifically for graphene, we have $h_{1 {\bs k}} = \hbar v_F k_x$, $h_{2 {\bs k}} = \hbar v_F k_y$ and $h_{3 {\bs k}} = 0$. Therefore, the optical matrix element for the low energy Hamiltonian of graphene in Eq.~\eqref{Gra_low} is,
\be 
{\bf M}_{\bs k}^{vc} = i e v_F\left(\sin\phi_{\bs k}, -\cos\phi_{\bs k},0\right),
\ee
where the azimuthal angle is given by $\phi_{\bs k} = \tan^{-1}(k_y/k_x)$.
%
Therefore, we have, ${\bf E}_p\cdot{\bf M}^{vc}_{\bs k} = -{\bf E}_p\cdot{\bf M}^{cv}_{\bs k}= iev_F|{\bf E}_p|\sin(\phi_{\bs k}-\Theta_p)$.
This leads to 
\be\label{n0_dimless}
\frac{n^{(0)}_{\bs k}}{n^{\rm eq}_{\bs k}} = \left[1 + \zeta^2\frac{\tilde{\gamma}_2^2}{\tilde{\omega}^2_{\bs k}}\frac{\sin^2\left(\phi_{\bs k}-\Theta_p\right)}{(\tilde{\omega}_{\bs k}-1)^2 + \tilde{\gamma}_2^2}\right]~.
\ee
where $\zeta = ev_F|{\bf E}_p|/(\hbar\omega_p\sqrt{\gamma_1\gamma_2})$, $\tilde\omega_{\bs k} = \omega_{\bs k}/
\omega_p$ and $\tilde\gamma_{2} = \gamma_{2}/\omega_p$. Accordingly we obtain, 
\be\label{s0k_dimless}
\sigma^{(0)}_{xx\bs k} = \frac{\sigma_0}{\pi^2}\frac{in^{(0)}_{\bs k}}{\tilde{\omega}_{\bs k}}\left[\frac{\sin^2\phi_{\bs k}}{\tilde{\omega}_{\bs k} - 1 - i\tilde{\gamma}_2}\right],
\ee
The total conductivity is obtained  by summing the momentum resolved conductivity over the BZ, 
\be
\sigma^{(0)}_{xx} = \int \tilde{\omega}_{\bs k}d\tilde{\omega}_{\bs k}d\phi_{\bs k}\sigma^{(0)}_{xx\bs k}~.
\ee
Similarly we obtain, 
\be\label{s1k_dimless}
\sigma^{(1)}_{xx\bs k} = \frac{\sigma_0}{\pi^2}\frac{in^{(0)}_{\bs k}}{\tilde{\omega}_{\bs k}}\left[\frac{\sin^2\phi_{\bs k}\left(1 + \tilde{\xi}_{\bf k}\sin^2\left(\phi_{\bs k}-\Theta_p\right)\right)}{\tilde{\omega}_{\bs k} - (1+\tilde{\omega}_1) - i\tilde{\gamma}_2}\right].
\ee
Here, we have defined 
\be
\tilde{\xi}_{\bf k} = \frac{\zeta^2\tilde{\gamma_1}\tilde{\gamma_2}\tilde{\mathcal P}_{\bs k}}{2\tilde{\omega}^2_{\bs k}(\tilde{\omega}_1 + i\tilde{\gamma}_1) + \zeta^2\tilde{\gamma_1}\tilde{\gamma_2}\sin^2\left(\phi_{\bs k}-\Theta_p\right)\tilde{\mathcal Q}_{\bs k}},
\ee
$\tilde{\mathcal P}_{\bs k} = {\mathcal P}_{\bs k}/\omega_p$ and $\tilde{\mathcal Q}_{\bs k} = {\mathcal Q}_{\bs k}/\omega_p$. 
Finally we have 
\be\label{smin1k_dimless}
\sigma^{(-1)}_{xx\bs k} = \frac{\sigma_0}{\pi^2}\frac{in^{(0)}_{\bs k}}{\tilde{\omega}_{\bs k}}\left[\frac{\tilde{\xi}^*_{\bf k}\sin^2\phi_{\bs k} \sin^2\left(\phi_{\bs k}-\Theta_p\right)}{\tilde{\omega}_{\bs k} - (1-\tilde{\omega}_1) - i\tilde{\gamma}_2}\right].
\ee
The dependence of the longitudinal conductivities defined above, on the optical field strength of the pump and its polarization dependence is shown in Fig.~\ref{conductivity}.

\begin{figure}[t]
\includegraphics[width = \linewidth]{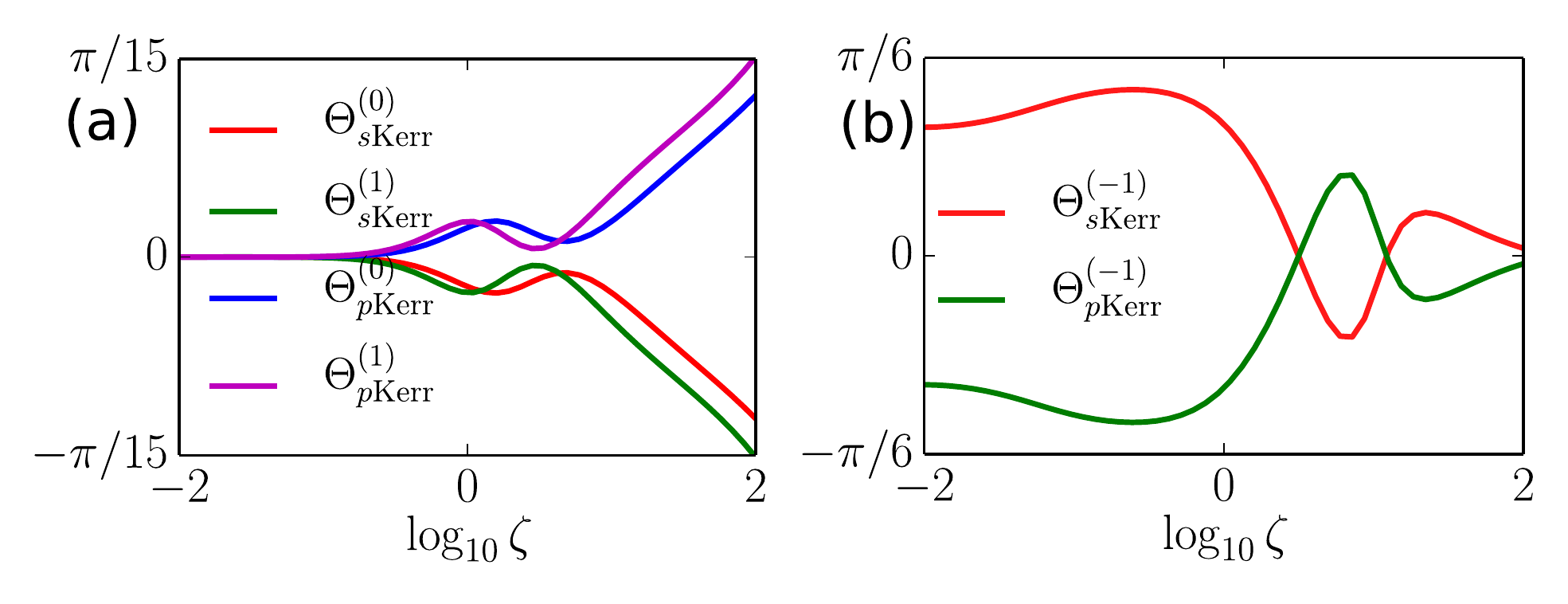}
\caption{Kerr angle for $s$ and $p$ component of pump, probe and the new sideband, as a function of $\log_{10}\zeta$. 
Other parameters are same as those of Fig. \ref{pop_inv}.
\label{Kerr_field}}
\end{figure}
\begin{figure}[t]
\includegraphics[width = \linewidth]{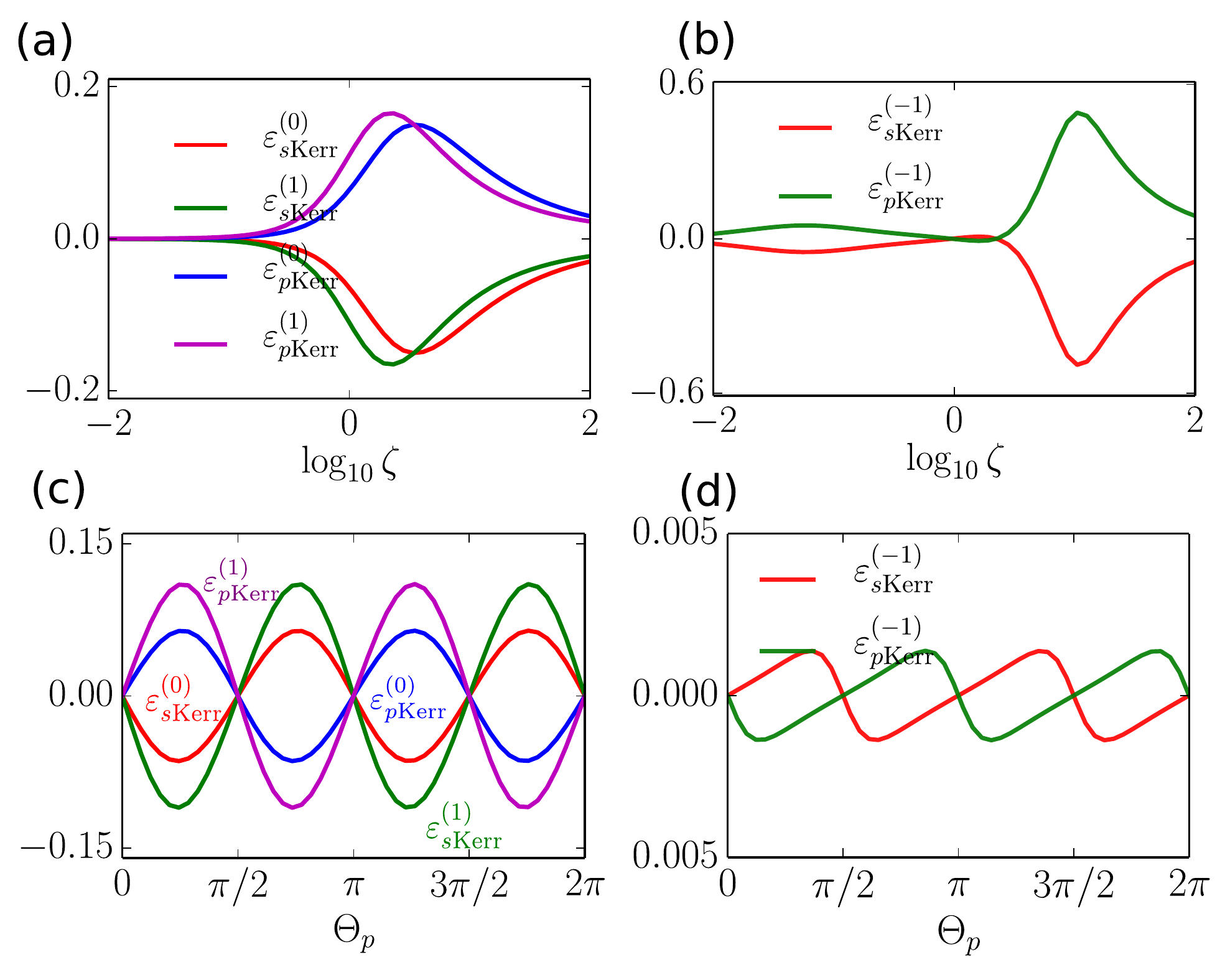}
\caption{Ellipticity for $s$ and $p$ component as a function of $\log_{10}\zeta$ for (a) the pump and probe frequencies, and (b) the new sideband. Dependence of ellipticity for $s$ and $p$ component as a function of $\Theta_p$ for (c) the pump and probe frequencies, and (d) the new sideband. Here $\Theta_p = \pi/4$ for (a) and (b) and $\zeta = 1$ for (c) and (d). Other parameters are same as those of Fig. \ref{pop_inv}.
\label{Ellipso}}
\end{figure}

In the non-linear response regime of $\zeta\geq1$, we also find the transverse optical conductivity ($\sigma_{xy}$) to be finite depending on the value of $\Theta_p$. 
The expressions for $\sigma_{xy\bs k}$ for pump, probe and the new sideband can be obtained directly from Eq.~\eqref{s0k_dimless}, Eq.~\eqref{s1k_dimless} and Eq.~\eqref{smin1k_dimless}, respectively, by replacing $\sin^2{\phi_{\bs k}}\to -\sin{\phi_{\bs k}}\cos{\phi_{\bs k}}$. The pump field intensity and polarization dependence of the transverse optical conductivity for graphene is shown in Fig.~\ref{transverse}. Note that these are in general smaller than the corresponding longitudinal counterparts. 

The presence of a finite optical conductivity also leads to polarization rotation in the reflected and transmitted optical beams. In particular, for the reflected beam, the Kerr angle is given by 
given by \cite{Yoshino:13,PhysRevB.97.205420},
\be
\Theta^{(\lambda)}_{s/p{\rm Kerr}} = \frac{1}{2}\tan^{-1}\left(\frac{2{{\rm Re}[\chi^{(\lambda)}_{s/p{\rm Kerr}}]}}{1 - |\chi^{(\lambda)}_{s/p{\rm Kerr}}|^2}\right).
\ee 
Here, $\lambda$ takes value $0$, $1$ and $-1$ for pump, probe and the new sideband frequencies, respectively. The explicit expressions for the $\chi_{s/p{\rm Kerr}}$ are given in Table I of the main manuscript. The dependence of the polarization rotation angle on the intensity of the pump beam is shown in Fig.~\ref{Kerr_field}. Evidently, the polarization angle of the newly generated optical sideband is significantly larger than the polarization rotation of the pump and probe fields. 

The corresponding ellipticity of the reflected beam is expressed as, 
\be
\varepsilon^{(\lambda)}_{s/p{\rm Kerr}} = \tan\left[\frac{1}{2}\sin^{-1}\left(\frac{2{{\rm Im}[\chi^{(\lambda)}_{s/p{\rm Kerr}}]}}{1 + |\chi^{(\lambda)}_{s/p{\rm Kerr}}|^2}\right)\right].
\ee 
The pump field and polarization angle dependence of the ellipticity of the reflected optical fields at pump, probe and sideband frequencies is shown in Fig.~\ref{Ellipso}.

\bibliography{refs_sideband}

\begin{thebibliography}{44}%
\makeatletter
\providecommand \@ifxundefined [1]{%
 \@ifx{#1\undefined}
}%
\providecommand \@ifnum [1]{%
 \ifnum #1\expandafter \@firstoftwo
 \else \expandafter \@secondoftwo
 \fi
}%
\providecommand \@ifx [1]{%
 \ifx #1\expandafter \@firstoftwo
 \else \expandafter \@secondoftwo
 \fi
}%
\providecommand \natexlab [1]{#1}%
\providecommand \enquote  [1]{``#1''}%
\providecommand \bibnamefont  [1]{#1}%
\providecommand \bibfnamefont [1]{#1}%
\providecommand \citenamefont [1]{#1}%
\providecommand \href@noop [0]{\@secondoftwo}%
\providecommand \href [0]{\begingroup \@sanitize@url \@href}%
\providecommand \@href[1]{\@@startlink{#1}\@@href}%
\providecommand \@@href[1]{\endgroup#1\@@endlink}%
\providecommand \@sanitize@url [0]{\catcode `\\12\catcode `\$12\catcode
  `\&12\catcode `\#12\catcode `\^12\catcode `\_12\catcode `\%12\relax}%
\providecommand \@@startlink[1]{}%
\providecommand \@@endlink[0]{}%
\providecommand \url  [0]{\begingroup\@sanitize@url \@url }%
\providecommand \@url [1]{\endgroup\@href {#1}{\urlprefix }}%
\providecommand \urlprefix  [0]{URL }%
\providecommand \Eprint [0]{\href }%
\providecommand \doibase [0]{http://dx.doi.org/}%
\providecommand \selectlanguage [0]{\@gobble}%
\providecommand \bibinfo  [0]{\@secondoftwo}%
\providecommand \bibfield  [0]{\@secondoftwo}%
\providecommand \translation [1]{[#1]}%
\providecommand \BibitemOpen [0]{}%
\providecommand \bibitemStop [0]{}%
\providecommand \bibitemNoStop [0]{.\EOS\space}%
\providecommand \EOS [0]{\spacefactor3000\relax}%
\providecommand \BibitemShut  [1]{\csname bibitem#1\endcsname}%
\let\auto@bib@innerbib\@empty
\bibitem [{\citenamefont {Hafez}\ \emph {et~al.}(2018)\citenamefont {Hafez},
  \citenamefont {Kovalev}, \citenamefont {Deinert}, \citenamefont {Mics},
  \citenamefont {Green}, \citenamefont {Awari}, \citenamefont {Chen},
  \citenamefont {Germanskiy}, \citenamefont {Lehnert}, \citenamefont
  {Teichert}, \citenamefont {Wang}, \citenamefont {Tielrooij}, \citenamefont
  {Liu}, \citenamefont {Chen}, \citenamefont {Narita}, \citenamefont
  {M{\"u}llen}, \citenamefont {Bonn}, \citenamefont {Gensch},\ and\
  \citenamefont {Turchinovich}}]{Hafez2018}%
  \BibitemOpen
  \bibfield  {author} {\bibinfo {author} {\bibfnamefont {Hassan~A.}\
  \bibnamefont {Hafez}}, \bibinfo {author} {\bibfnamefont {Sergey}\
  \bibnamefont {Kovalev}}, \bibinfo {author} {\bibfnamefont {Jan-Christoph}\
  \bibnamefont {Deinert}}, \bibinfo {author} {\bibfnamefont {Zolt{\'a}n}\
  \bibnamefont {Mics}}, \bibinfo {author} {\bibfnamefont {Bertram}\
  \bibnamefont {Green}}, \bibinfo {author} {\bibfnamefont {Nilesh}\
  \bibnamefont {Awari}}, \bibinfo {author} {\bibfnamefont {Min}\ \bibnamefont
  {Chen}}, \bibinfo {author} {\bibfnamefont {Semyon}\ \bibnamefont
  {Germanskiy}}, \bibinfo {author} {\bibfnamefont {Ulf}\ \bibnamefont
  {Lehnert}}, \bibinfo {author} {\bibfnamefont {Jochen}\ \bibnamefont
  {Teichert}}, \bibinfo {author} {\bibfnamefont {Zhe}\ \bibnamefont {Wang}},
  \bibinfo {author} {\bibfnamefont {Klaas-Jan}\ \bibnamefont {Tielrooij}},
  \bibinfo {author} {\bibfnamefont {Zhaoyang}\ \bibnamefont {Liu}}, \bibinfo
  {author} {\bibfnamefont {Zongping}\ \bibnamefont {Chen}}, \bibinfo {author}
  {\bibfnamefont {Akimitsu}\ \bibnamefont {Narita}}, \bibinfo {author}
  {\bibfnamefont {Klaus}\ \bibnamefont {M{\"u}llen}}, \bibinfo {author}
  {\bibfnamefont {Mischa}\ \bibnamefont {Bonn}}, \bibinfo {author}
  {\bibfnamefont {Michael}\ \bibnamefont {Gensch}}, \ and\ \bibinfo {author}
  {\bibfnamefont {Dmitry}\ \bibnamefont {Turchinovich}},\ }\bibfield  {title}
  {\enquote {\bibinfo {title} {Extremely efficient terahertz high-harmonic
  generation in graphene by hot dirac fermions},}\ }\href {\doibase
  10.1038/s41586-018-0508-1} {\bibfield  {journal} {\bibinfo  {journal}
  {Nature}\ }\textbf {\bibinfo {volume} {561}},\ \bibinfo {pages} {507--511}
  (\bibinfo {year} {2018})}\BibitemShut {NoStop}%
\bibitem [{\citenamefont {Yoshikawa}\ \emph {et~al.}(2017)\citenamefont
  {Yoshikawa}, \citenamefont {Tamaya},\ and\ \citenamefont
  {Tanaka}}]{Yoshikawa2017}%
  \BibitemOpen
  \bibfield  {author} {\bibinfo {author} {\bibfnamefont {Naotaka}\ \bibnamefont
  {Yoshikawa}}, \bibinfo {author} {\bibfnamefont {Tomohiro}\ \bibnamefont
  {Tamaya}}, \ and\ \bibinfo {author} {\bibfnamefont {Koichiro}\ \bibnamefont
  {Tanaka}},\ }\bibfield  {title} {\enquote {\bibinfo {title} {{Optics:
  High-harmonic generation in graphene enhanced by elliptically polarized light
  excitation}},}\ }\href {\doibase 10.1126/science.aam8861} {\bibfield
  {journal} {\bibinfo  {journal} {Science}\ }\textbf {\bibinfo {volume}
  {356}},\ \bibinfo {pages} {736--738} (\bibinfo {year} {2017})}\BibitemShut
  {NoStop}%
\bibitem [{\citenamefont {Prechtel}\ \emph {et~al.}(2012)\citenamefont
  {Prechtel}, \citenamefont {Song}, \citenamefont {Schuh}, \citenamefont
  {Ajayan}, \citenamefont {Wegscheider},\ and\ \citenamefont
  {Holleitner}}]{Prechtel2012}%
  \BibitemOpen
  \bibfield  {author} {\bibinfo {author} {\bibfnamefont {Leonhard}\
  \bibnamefont {Prechtel}}, \bibinfo {author} {\bibfnamefont {Li}~\bibnamefont
  {Song}}, \bibinfo {author} {\bibfnamefont {Dieter}\ \bibnamefont {Schuh}},
  \bibinfo {author} {\bibfnamefont {Pulickel}\ \bibnamefont {Ajayan}}, \bibinfo
  {author} {\bibfnamefont {Werner}\ \bibnamefont {Wegscheider}}, \ and\
  \bibinfo {author} {\bibfnamefont {Alexander~W.}\ \bibnamefont {Holleitner}},\
  }\bibfield  {title} {\enquote {\bibinfo {title} {Time-resolved ultrafast
  photocurrents and terahertz generation in freely suspended graphene},}\
  }\href {http://dx.doi.org/10.1038/ncomms1656} {\bibfield  {journal} {\bibinfo
   {journal} {Nature Communications}\ }\textbf {\bibinfo {volume} {3}},\
  \bibinfo {pages} {646} (\bibinfo {year} {2012})}\BibitemShut {NoStop}%
\bibitem [{\citenamefont {Jiang}\ \emph {et~al.}(2018)\citenamefont {Jiang},
  \citenamefont {Huang}, \citenamefont {Cheng}, \citenamefont {Fan},
  \citenamefont {Zhang}, \citenamefont {Shan}, \citenamefont {Yi},
  \citenamefont {Dai}, \citenamefont {Shi}, \citenamefont {Liu}, \citenamefont
  {Zeng}, \citenamefont {Zi}, \citenamefont {Sipe}, \citenamefont {Shen},
  \citenamefont {Liu},\ and\ \citenamefont {Wu}}]{Jiang2018}%
  \BibitemOpen
  \bibfield  {author} {\bibinfo {author} {\bibfnamefont {Tao}\ \bibnamefont
  {Jiang}}, \bibinfo {author} {\bibfnamefont {Di}~\bibnamefont {Huang}},
  \bibinfo {author} {\bibfnamefont {Jinluo}\ \bibnamefont {Cheng}}, \bibinfo
  {author} {\bibfnamefont {Xiaodong}\ \bibnamefont {Fan}}, \bibinfo {author}
  {\bibfnamefont {Zhihong}\ \bibnamefont {Zhang}}, \bibinfo {author}
  {\bibfnamefont {Yuwei}\ \bibnamefont {Shan}}, \bibinfo {author}
  {\bibfnamefont {Yangfan}\ \bibnamefont {Yi}}, \bibinfo {author}
  {\bibfnamefont {Yunyun}\ \bibnamefont {Dai}}, \bibinfo {author}
  {\bibfnamefont {Lei}\ \bibnamefont {Shi}}, \bibinfo {author} {\bibfnamefont
  {Kaihui}\ \bibnamefont {Liu}}, \bibinfo {author} {\bibfnamefont {Changgan}\
  \bibnamefont {Zeng}}, \bibinfo {author} {\bibfnamefont {Jian}\ \bibnamefont
  {Zi}}, \bibinfo {author} {\bibfnamefont {J.~E.}\ \bibnamefont {Sipe}},
  \bibinfo {author} {\bibfnamefont {Yuen~Ron}\ \bibnamefont {Shen}}, \bibinfo
  {author} {\bibfnamefont {Wei~Tao}\ \bibnamefont {Liu}}, \ and\ \bibinfo
  {author} {\bibfnamefont {Shiwei}\ \bibnamefont {Wu}},\ }\bibfield  {title}
  {\enquote {\bibinfo {title} {{Gate-tunable third-order nonlinear optical
  response of massless Dirac fermions in graphene}},}\ }\href {\doibase
  10.1038/s41566-018-0175-7} {\bibfield  {journal} {\bibinfo  {journal} {Nature
  Photonics}\ }\textbf {\bibinfo {volume} {12}},\ \bibinfo {pages} {430--436}
  (\bibinfo {year} {2018})}\BibitemShut {NoStop}%
\bibitem [{\citenamefont {Gu}\ \emph {et~al.}(2012)\citenamefont {Gu},
  \citenamefont {Petrone}, \citenamefont {McMillan}, \citenamefont {{Van Der
  Zande}}, \citenamefont {Yu}, \citenamefont {Lo}, \citenamefont {Kwong},
  \citenamefont {Hone},\ and\ \citenamefont {Wong}}]{Gu2012}%
  \BibitemOpen
  \bibfield  {author} {\bibinfo {author} {\bibfnamefont {T.}~\bibnamefont
  {Gu}}, \bibinfo {author} {\bibfnamefont {N.}~\bibnamefont {Petrone}},
  \bibinfo {author} {\bibfnamefont {J.~F.}\ \bibnamefont {McMillan}}, \bibinfo
  {author} {\bibfnamefont {A.}~\bibnamefont {{Van Der Zande}}}, \bibinfo
  {author} {\bibfnamefont {M.}~\bibnamefont {Yu}}, \bibinfo {author}
  {\bibfnamefont {G.~Q.}\ \bibnamefont {Lo}}, \bibinfo {author} {\bibfnamefont
  {D.~L.}\ \bibnamefont {Kwong}}, \bibinfo {author} {\bibfnamefont
  {J.}~\bibnamefont {Hone}}, \ and\ \bibinfo {author} {\bibfnamefont {C.~W.}\
  \bibnamefont {Wong}},\ }\bibfield  {title} {\enquote {\bibinfo {title}
  {{Regenerative oscillation and four-wave mixing in graphene
  optoelectronics}},}\ }\href {\doibase 10.1038/nphoton.2012.147} {\bibfield
  {journal} {\bibinfo  {journal} {Nature Photonics}\ }\textbf {\bibinfo
  {volume} {6}},\ \bibinfo {pages} {554--559} (\bibinfo {year}
  {2012})}\BibitemShut {NoStop}%
\bibitem [{\citenamefont {Hendry}\ \emph {et~al.}(2010)\citenamefont {Hendry},
  \citenamefont {Hale}, \citenamefont {Moger}, \citenamefont {Savchenko},\ and\
  \citenamefont {Mikhailov}}]{Hendry}%
  \BibitemOpen
  \bibfield  {author} {\bibinfo {author} {\bibfnamefont {E.}~\bibnamefont
  {Hendry}}, \bibinfo {author} {\bibfnamefont {P.~J.}\ \bibnamefont {Hale}},
  \bibinfo {author} {\bibfnamefont {J.}~\bibnamefont {Moger}}, \bibinfo
  {author} {\bibfnamefont {A.~K.}\ \bibnamefont {Savchenko}}, \ and\ \bibinfo
  {author} {\bibfnamefont {S.~A.}\ \bibnamefont {Mikhailov}},\ }\bibfield
  {title} {\enquote {\bibinfo {title} {Coherent nonlinear optical response of
  graphene},}\ }\href {\doibase 10.1103/PhysRevLett.105.097401} {\bibfield
  {journal} {\bibinfo  {journal} {Phys. Rev. Lett.}\ }\textbf {\bibinfo
  {volume} {105}},\ \bibinfo {pages} {097401} (\bibinfo {year}
  {2010})}\BibitemShut {NoStop}%
\bibitem [{\citenamefont {Yang}\ \emph {et~al.}(2011)\citenamefont {Yang},
  \citenamefont {Feng}, \citenamefont {Wang}, \citenamefont {Huang},
  \citenamefont {Chen}, \citenamefont {Wee},\ and\ \citenamefont {Ji}}]{Yang}%
  \BibitemOpen
  \bibfield  {author} {\bibinfo {author} {\bibfnamefont {Hongzhi}\ \bibnamefont
  {Yang}}, \bibinfo {author} {\bibfnamefont {Xiaobo}\ \bibnamefont {Feng}},
  \bibinfo {author} {\bibfnamefont {Qian}\ \bibnamefont {Wang}}, \bibinfo
  {author} {\bibfnamefont {Han}\ \bibnamefont {Huang}}, \bibinfo {author}
  {\bibfnamefont {Wei}\ \bibnamefont {Chen}}, \bibinfo {author} {\bibfnamefont
  {Andrew T.~S.}\ \bibnamefont {Wee}}, \ and\ \bibinfo {author} {\bibfnamefont
  {Wei}\ \bibnamefont {Ji}},\ }\bibfield  {title} {\enquote {\bibinfo {title}
  {Giant two-photon absorption in bilayer graphene},}\ }\href {\doibase
  10.1021/nl200587h} {\bibfield  {journal} {\bibinfo  {journal} {Nano Letters}\
  }\textbf {\bibinfo {volume} {11}},\ \bibinfo {pages} {2622--2627} (\bibinfo
  {year} {2011})}\BibitemShut {NoStop}%
\bibitem [{\citenamefont {Zhang}\ \emph {et~al.}(2012)\citenamefont {Zhang},
  \citenamefont {Virally}, \citenamefont {Bao}, \citenamefont {Ping},
  \citenamefont {Massar}, \citenamefont {Godbout},\ and\ \citenamefont
  {Kockaert}}]{Zhang}%
  \BibitemOpen
  \bibfield  {author} {\bibinfo {author} {\bibfnamefont {Han}\ \bibnamefont
  {Zhang}}, \bibinfo {author} {\bibfnamefont {St\'{e}phane}\ \bibnamefont
  {Virally}}, \bibinfo {author} {\bibfnamefont {Qiaoliang}\ \bibnamefont
  {Bao}}, \bibinfo {author} {\bibfnamefont {Loh~Kian}\ \bibnamefont {Ping}},
  \bibinfo {author} {\bibfnamefont {Serge}\ \bibnamefont {Massar}}, \bibinfo
  {author} {\bibfnamefont {Nicolas}\ \bibnamefont {Godbout}}, \ and\ \bibinfo
  {author} {\bibfnamefont {Pascal}\ \bibnamefont {Kockaert}},\ }\bibfield
  {title} {\enquote {\bibinfo {title} {Z-scan measurement of the nonlinear
  refractive index of graphene},}\ }\href {\doibase 10.1364/OL.37.001856}
  {\bibfield  {journal} {\bibinfo  {journal} {Opt. Lett.}\ }\textbf {\bibinfo
  {volume} {37}},\ \bibinfo {pages} {1856--1858} (\bibinfo {year}
  {2012})}\BibitemShut {NoStop}%
\bibitem [{\citenamefont {Kumar}\ \emph {et~al.}(2013)\citenamefont {Kumar},
  \citenamefont {Kumar}, \citenamefont {Gerstenkorn}, \citenamefont {Wang},
  \citenamefont {Chiu}, \citenamefont {Smirl},\ and\ \citenamefont
  {Zhao}}]{Kumar}%
  \BibitemOpen
  \bibfield  {author} {\bibinfo {author} {\bibfnamefont {Nardeep}\ \bibnamefont
  {Kumar}}, \bibinfo {author} {\bibfnamefont {Jatinder}\ \bibnamefont {Kumar}},
  \bibinfo {author} {\bibfnamefont {Chris}\ \bibnamefont {Gerstenkorn}},
  \bibinfo {author} {\bibfnamefont {Rui}\ \bibnamefont {Wang}}, \bibinfo
  {author} {\bibfnamefont {Hsin-Ying}\ \bibnamefont {Chiu}}, \bibinfo {author}
  {\bibfnamefont {Arthur~L.}\ \bibnamefont {Smirl}}, \ and\ \bibinfo {author}
  {\bibfnamefont {Hui}\ \bibnamefont {Zhao}},\ }\bibfield  {title} {\enquote
  {\bibinfo {title} {Third harmonic generation in graphene and few-layer
  graphite films},}\ }\href {\doibase 10.1103/PhysRevB.87.121406} {\bibfield
  {journal} {\bibinfo  {journal} {Phys. Rev. B}\ }\textbf {\bibinfo {volume}
  {87}},\ \bibinfo {pages} {121406} (\bibinfo {year} {2013})}\BibitemShut
  {NoStop}%
\bibitem [{\citenamefont {Hong}\ \emph {et~al.}(2013)\citenamefont {Hong},
  \citenamefont {Dadap}, \citenamefont {Petrone}, \citenamefont {Yeh},
  \citenamefont {Hone},\ and\ \citenamefont {Osgood}}]{Hong}%
  \BibitemOpen
  \bibfield  {author} {\bibinfo {author} {\bibfnamefont {Sung-Young}\
  \bibnamefont {Hong}}, \bibinfo {author} {\bibfnamefont {Jerry~I.}\
  \bibnamefont {Dadap}}, \bibinfo {author} {\bibfnamefont {Nicholas}\
  \bibnamefont {Petrone}}, \bibinfo {author} {\bibfnamefont {Po-Chun}\
  \bibnamefont {Yeh}}, \bibinfo {author} {\bibfnamefont {James}\ \bibnamefont
  {Hone}}, \ and\ \bibinfo {author} {\bibfnamefont {Richard~M.}\ \bibnamefont
  {Osgood}},\ }\bibfield  {title} {\enquote {\bibinfo {title} {Optical
  third-harmonic generation in graphene},}\ }\href {\doibase
  10.1103/PhysRevX.3.021014} {\bibfield  {journal} {\bibinfo  {journal} {Phys.
  Rev. X}\ }\textbf {\bibinfo {volume} {3}},\ \bibinfo {pages} {021014}
  (\bibinfo {year} {2013})}\BibitemShut {NoStop}%
\bibitem [{\citenamefont {Mikhailov}\ and\ \citenamefont
  {Ziegler}(2008)}]{JPCC_Mikhailov}%
  \BibitemOpen
  \bibfield  {author} {\bibinfo {author} {\bibfnamefont {S.~A.}\ \bibnamefont
  {Mikhailov}}\ and\ \bibinfo {author} {\bibfnamefont {K.}~\bibnamefont
  {Ziegler}},\ }\bibfield  {title} {\enquote {\bibinfo {title} {Nonlinear
  electromagnetic response of graphene: frequency multiplication and the
  self-consistent-field effects},}\ }\href
  {http://stacks.iop.org/0953-8984/20/i=38/a=384204} {\bibfield  {journal}
  {\bibinfo  {journal} {Journal of Physics: Condensed Matter}\ }\textbf
  {\bibinfo {volume} {20}},\ \bibinfo {pages} {384204} (\bibinfo {year}
  {2008})}\BibitemShut {NoStop}%
\bibitem [{\citenamefont {Glazov}\ and\ \citenamefont
  {Ganichev}(2014)}]{Glazov2014}%
  \BibitemOpen
  \bibfield  {author} {\bibinfo {author} {\bibfnamefont {M.~M.}\ \bibnamefont
  {Glazov}}\ and\ \bibinfo {author} {\bibfnamefont {S.~D.}\ \bibnamefont
  {Ganichev}},\ }\bibfield  {title} {\enquote {\bibinfo {title} {{High
  frequency electric field induced nonlinear effects in graphene}},}\ }\href
  {\doibase 10.1016/j.physrep.2013.10.003} {\bibfield  {journal} {\bibinfo
  {journal} {Physics Reports}\ }\textbf {\bibinfo {volume} {535}},\ \bibinfo
  {pages} {101--138} (\bibinfo {year} {2014})}\BibitemShut {NoStop}%
\bibitem [{\citenamefont {Cheng}\ \emph {et~al.}(2014)\citenamefont {Cheng},
  \citenamefont {Vermeulen},\ and\ \citenamefont {Sipe}}]{Cheng2014}%
  \BibitemOpen
  \bibfield  {author} {\bibinfo {author} {\bibfnamefont {J~L}\ \bibnamefont
  {Cheng}}, \bibinfo {author} {\bibfnamefont {N}~\bibnamefont {Vermeulen}}, \
  and\ \bibinfo {author} {\bibfnamefont {J~E}\ \bibnamefont {Sipe}},\
  }\bibfield  {title} {\enquote {\bibinfo {title} {Third order optical
  nonlinearity of graphene},}\ }\href
  {http://stacks.iop.org/1367-2630/16/i=5/a=053014} {\bibfield  {journal}
  {\bibinfo  {journal} {New Journal of Physics}\ }\textbf {\bibinfo {volume}
  {16}},\ \bibinfo {pages} {053014} (\bibinfo {year} {2014})}\BibitemShut
  {NoStop}%
\bibitem [{\citenamefont {Al-Naib}\ \emph {et~al.}(2014)\citenamefont
  {Al-Naib}, \citenamefont {Sipe},\ and\ \citenamefont
  {Dignam}}]{PhysRevB.90.245423}%
  \BibitemOpen
  \bibfield  {author} {\bibinfo {author} {\bibfnamefont {Ibraheem}\
  \bibnamefont {Al-Naib}}, \bibinfo {author} {\bibfnamefont {J.~E.}\
  \bibnamefont {Sipe}}, \ and\ \bibinfo {author} {\bibfnamefont {Marc~M.}\
  \bibnamefont {Dignam}},\ }\bibfield  {title} {\enquote {\bibinfo {title}
  {High harmonic generation in undoped graphene: Interplay of inter- and
  intraband dynamics},}\ }\href {\doibase 10.1103/PhysRevB.90.245423}
  {\bibfield  {journal} {\bibinfo  {journal} {Phys. Rev. B}\ }\textbf {\bibinfo
  {volume} {90}},\ \bibinfo {pages} {245423} (\bibinfo {year}
  {2014})}\BibitemShut {NoStop}%
\bibitem [{\citenamefont {Tamaya}\ \emph
  {et~al.}(2016{\natexlab{a}})\citenamefont {Tamaya}, \citenamefont {Ishikawa},
  \citenamefont {Ogawa},\ and\ \citenamefont
  {Tanaka}}]{PhysRevLett.116.016601}%
  \BibitemOpen
  \bibfield  {author} {\bibinfo {author} {\bibfnamefont {T.}~\bibnamefont
  {Tamaya}}, \bibinfo {author} {\bibfnamefont {A.}~\bibnamefont {Ishikawa}},
  \bibinfo {author} {\bibfnamefont {T.}~\bibnamefont {Ogawa}}, \ and\ \bibinfo
  {author} {\bibfnamefont {K.}~\bibnamefont {Tanaka}},\ }\bibfield  {title}
  {\enquote {\bibinfo {title} {Diabatic mechanisms of higher-order harmonic
  generation in solid-state materials under high-intensity electric fields},}\
  }\href {\doibase 10.1103/PhysRevLett.116.016601} {\bibfield  {journal}
  {\bibinfo  {journal} {Phys. Rev. Lett.}\ }\textbf {\bibinfo {volume} {116}},\
  \bibinfo {pages} {016601} (\bibinfo {year} {2016}{\natexlab{a}})}\BibitemShut
  {NoStop}%
\bibitem [{\citenamefont {Cheng}\ \emph {et~al.}(2015)\citenamefont {Cheng},
  \citenamefont {Vermeulen},\ and\ \citenamefont {Sipe}}]{PhysRevB.91.235320}%
  \BibitemOpen
  \bibfield  {author} {\bibinfo {author} {\bibfnamefont {J.~L.}\ \bibnamefont
  {Cheng}}, \bibinfo {author} {\bibfnamefont {N.}~\bibnamefont {Vermeulen}}, \
  and\ \bibinfo {author} {\bibfnamefont {J.~E.}\ \bibnamefont {Sipe}},\
  }\bibfield  {title} {\enquote {\bibinfo {title} {Third-order nonlinearity of
  graphene: Effects of phenomenological relaxation and finite temperature},}\
  }\href {\doibase 10.1103/PhysRevB.91.235320} {\bibfield  {journal} {\bibinfo
  {journal} {Phys. Rev. B}\ }\textbf {\bibinfo {volume} {91}},\ \bibinfo
  {pages} {235320} (\bibinfo {year} {2015})}\BibitemShut {NoStop}%
\bibitem [{\citenamefont {Mikhailov}(2016)}]{PhysRevB.93.085403}%
  \BibitemOpen
  \bibfield  {author} {\bibinfo {author} {\bibfnamefont {S.~A.}\ \bibnamefont
  {Mikhailov}},\ }\bibfield  {title} {\enquote {\bibinfo {title} {Quantum
  theory of the third-order nonlinear electrodynamic effects of graphene},}\
  }\href {\doibase 10.1103/PhysRevB.93.085403} {\bibfield  {journal} {\bibinfo
  {journal} {Phys. Rev. B}\ }\textbf {\bibinfo {volume} {93}},\ \bibinfo
  {pages} {085403} (\bibinfo {year} {2016})}\BibitemShut {NoStop}%
\bibitem [{\citenamefont {Rostami}\ and\ \citenamefont
  {Polini}(2016)}]{PhysRevB.93.161411}%
  \BibitemOpen
  \bibfield  {author} {\bibinfo {author} {\bibfnamefont {Habib}\ \bibnamefont
  {Rostami}}\ and\ \bibinfo {author} {\bibfnamefont {Marco}\ \bibnamefont
  {Polini}},\ }\bibfield  {title} {\enquote {\bibinfo {title} {Theory of
  third-harmonic generation in graphene: A diagrammatic approach},}\ }\href
  {\doibase 10.1103/PhysRevB.93.161411} {\bibfield  {journal} {\bibinfo
  {journal} {Phys. Rev. B}\ }\textbf {\bibinfo {volume} {93}},\ \bibinfo
  {pages} {161411} (\bibinfo {year} {2016})}\BibitemShut {NoStop}%
\bibitem [{\citenamefont {Gullans}\ \emph {et~al.}(2013)\citenamefont
  {Gullans}, \citenamefont {Chang}, \citenamefont {Koppens}, \citenamefont
  {de~Abajo},\ and\ \citenamefont {Lukin}}]{Gullans2013}%
  \BibitemOpen
  \bibfield  {author} {\bibinfo {author} {\bibfnamefont {M.}~\bibnamefont
  {Gullans}}, \bibinfo {author} {\bibfnamefont {D.~E.}\ \bibnamefont {Chang}},
  \bibinfo {author} {\bibfnamefont {F.~H.~L.}\ \bibnamefont {Koppens}},
  \bibinfo {author} {\bibfnamefont {F.~J.~G.}\ \bibnamefont {de~Abajo}}, \ and\
  \bibinfo {author} {\bibfnamefont {M.~D.}\ \bibnamefont {Lukin}},\ }\bibfield
  {title} {\enquote {\bibinfo {title} {Single-photon nonlinear optics with
  graphene plasmons},}\ }\href {\doibase 10.1103/PhysRevLett.111.247401}
  {\bibfield  {journal} {\bibinfo  {journal} {Phys. Rev. Lett.}\ }\textbf
  {\bibinfo {volume} {111}},\ \bibinfo {pages} {247401} (\bibinfo {year}
  {2013})}\BibitemShut {NoStop}%
\bibitem [{\citenamefont {Yao}\ \emph {et~al.}(2014)\citenamefont {Yao},
  \citenamefont {Tokman},\ and\ \citenamefont {Belyanin}}]{Yao2014}%
  \BibitemOpen
  \bibfield  {author} {\bibinfo {author} {\bibfnamefont {Xianghan}\
  \bibnamefont {Yao}}, \bibinfo {author} {\bibfnamefont {Mikhail}\ \bibnamefont
  {Tokman}}, \ and\ \bibinfo {author} {\bibfnamefont {Alexey}\ \bibnamefont
  {Belyanin}},\ }\bibfield  {title} {\enquote {\bibinfo {title} {Efficient
  nonlinear generation of thz plasmons in graphene and topological
  insulators},}\ }\href {\doibase 10.1103/PhysRevLett.112.055501} {\bibfield
  {journal} {\bibinfo  {journal} {Phys. Rev. Lett.}\ }\textbf {\bibinfo
  {volume} {112}},\ \bibinfo {pages} {055501} (\bibinfo {year}
  {2014})}\BibitemShut {NoStop}%
\bibitem [{\citenamefont {Crosse}\ \emph {et~al.}(2014)\citenamefont {Crosse},
  \citenamefont {Xu}, \citenamefont {Sherwin},\ and\ \citenamefont
  {Liu}}]{Crosse2014}%
  \BibitemOpen
  \bibfield  {author} {\bibinfo {author} {\bibfnamefont {J.~A.}\ \bibnamefont
  {Crosse}}, \bibinfo {author} {\bibfnamefont {Xiaodong}\ \bibnamefont {Xu}},
  \bibinfo {author} {\bibfnamefont {Mark~S.}\ \bibnamefont {Sherwin}}, \ and\
  \bibinfo {author} {\bibfnamefont {R.~B.}\ \bibnamefont {Liu}},\ }\bibfield
  {title} {\enquote {\bibinfo {title} {Theory of low-power ultra-broadband
  terahertz sideband generation in bi-layer graphene},}\ }\href
  {http://dx.doi.org/10.1038/ncomms5854} {\bibfield  {journal} {\bibinfo
  {journal} {Nature Communications}\ }\textbf {\bibinfo {volume} {5}},\
  \bibinfo {pages} {4854} (\bibinfo {year} {2014})}\BibitemShut {NoStop}%
\bibitem [{\citenamefont {Cox}\ \emph {et~al.}(2017)\citenamefont {Cox},
  \citenamefont {Marini},\ and\ \citenamefont {de~Abajo}}]{Cox2017}%
  \BibitemOpen
  \bibfield  {author} {\bibinfo {author} {\bibfnamefont {Joel~D.}\ \bibnamefont
  {Cox}}, \bibinfo {author} {\bibfnamefont {Andrea}\ \bibnamefont {Marini}}, \
  and\ \bibinfo {author} {\bibfnamefont {F.~Javier~Garcia}\ \bibnamefont
  {de~Abajo}},\ }\bibfield  {title} {\enquote {\bibinfo {title}
  {Plasmon-assisted high-harmonic generation in graphene},}\ }\href
  {http://dx.doi.org/10.1038/ncomms14380} {\bibfield  {journal} {\bibinfo
  {journal} {Nature Communications}\ }\textbf {\bibinfo {volume} {8}},\
  \bibinfo {pages} {14380} (\bibinfo {year} {2017})}\BibitemShut {NoStop}%
\bibitem [{\citenamefont {Bonaccorso}\ \emph {et~al.}(2010)\citenamefont
  {Bonaccorso}, \citenamefont {Sun}, \citenamefont {Hasan},\ and\ \citenamefont
  {Ferrari}}]{Bonaccorso2010}%
  \BibitemOpen
  \bibfield  {author} {\bibinfo {author} {\bibfnamefont {F.}~\bibnamefont
  {Bonaccorso}}, \bibinfo {author} {\bibfnamefont {Z.}~\bibnamefont {Sun}},
  \bibinfo {author} {\bibfnamefont {T.}~\bibnamefont {Hasan}}, \ and\ \bibinfo
  {author} {\bibfnamefont {A.~C.}\ \bibnamefont {Ferrari}},\ }\bibfield
  {title} {\enquote {\bibinfo {title} {{Graphene photonics and
  optoelectronics}},}\ }\href {\doibase 10.1038/nphoton.2010.186} {\bibfield
  {journal} {\bibinfo  {journal} {Nature Photonics}\ }\textbf {\bibinfo
  {volume} {4}},\ \bibinfo {pages} {611--622} (\bibinfo {year}
  {2010})}\BibitemShut {NoStop}%
\bibitem [{\citenamefont {Bao}\ and\ \citenamefont {Loh}(2012)}]{Bao2012}%
  \BibitemOpen
  \bibfield  {author} {\bibinfo {author} {\bibfnamefont {Qiaoliang}\
  \bibnamefont {Bao}}\ and\ \bibinfo {author} {\bibfnamefont {Kian~Ping}\
  \bibnamefont {Loh}},\ }\bibfield  {title} {\enquote {\bibinfo {title}
  {{Graphene photonics, plasmonics, and broadband optoelectronic devices}},}\
  }\href {\doibase 10.1021/nn300989g} {\bibfield  {journal} {\bibinfo
  {journal} {ACS Nano}\ }\textbf {\bibinfo {volume} {6}},\ \bibinfo {pages}
  {3677--3694} (\bibinfo {year} {2012})}\BibitemShut {NoStop}%
\bibitem [{\citenamefont {Tamaya}\ \emph
  {et~al.}(2016{\natexlab{b}})\citenamefont {Tamaya}, \citenamefont {Ishikawa},
  \citenamefont {Ogawa},\ and\ \citenamefont {Tanaka}}]{PhysRevB.94.241107}%
  \BibitemOpen
  \bibfield  {author} {\bibinfo {author} {\bibfnamefont {T.}~\bibnamefont
  {Tamaya}}, \bibinfo {author} {\bibfnamefont {A.}~\bibnamefont {Ishikawa}},
  \bibinfo {author} {\bibfnamefont {T.}~\bibnamefont {Ogawa}}, \ and\ \bibinfo
  {author} {\bibfnamefont {K.}~\bibnamefont {Tanaka}},\ }\bibfield  {title}
  {\enquote {\bibinfo {title} {Higher-order harmonic generation caused by
  elliptically polarized electric fields in solid-state materials},}\ }\href
  {\doibase 10.1103/PhysRevB.94.241107} {\bibfield  {journal} {\bibinfo
  {journal} {Phys. Rev. B}\ }\textbf {\bibinfo {volume} {94}},\ \bibinfo
  {pages} {241107} (\bibinfo {year} {2016}{\natexlab{b}})}\BibitemShut
  {NoStop}%
\bibitem [{\citenamefont {Dawlaty}\ \emph {et~al.}(2008)\citenamefont
  {Dawlaty}, \citenamefont {Shivaraman}, \citenamefont {Chandrashekhar},
  \citenamefont {Rana},\ and\ \citenamefont {Spencer}}]{Dawlaty2008}%
  \BibitemOpen
  \bibfield  {author} {\bibinfo {author} {\bibfnamefont {Jahan~M.}\
  \bibnamefont {Dawlaty}}, \bibinfo {author} {\bibfnamefont {Shriram}\
  \bibnamefont {Shivaraman}}, \bibinfo {author} {\bibfnamefont {Mvs}\
  \bibnamefont {Chandrashekhar}}, \bibinfo {author} {\bibfnamefont {Farhan}\
  \bibnamefont {Rana}}, \ and\ \bibinfo {author} {\bibfnamefont {Michael~G.}\
  \bibnamefont {Spencer}},\ }\bibfield  {title} {\enquote {\bibinfo {title}
  {Measurement of ultrafast carrier dynamics in epitaxial graphene},}\ }\href
  {\doibase 10.1063/1.2837539} {\bibfield  {journal} {\bibinfo  {journal}
  {Applied Physics Letters}\ }\textbf {\bibinfo {volume} {92}},\ \bibinfo
  {pages} {042116} (\bibinfo {year} {2008})}\BibitemShut {NoStop}%
\bibitem [{\citenamefont {Johannsen}\ \emph {et~al.}(2013)\citenamefont
  {Johannsen}, \citenamefont {Ulstrup}, \citenamefont {Cilento}, \citenamefont
  {Crepaldi}, \citenamefont {Zacchigna}, \citenamefont {Cacho}, \citenamefont
  {Turcu}, \citenamefont {Springate}, \citenamefont {Fromm}, \citenamefont
  {Raidel}, \citenamefont {Seyller}, \citenamefont {Parmigiani}, \citenamefont
  {Grioni},\ and\ \citenamefont {Hofmann}}]{PhysRevLett.111.027403}%
  \BibitemOpen
  \bibfield  {author} {\bibinfo {author} {\bibfnamefont {Jens~Christian}\
  \bibnamefont {Johannsen}}, \bibinfo {author} {\bibfnamefont {S\o{}ren}\
  \bibnamefont {Ulstrup}}, \bibinfo {author} {\bibfnamefont {Federico}\
  \bibnamefont {Cilento}}, \bibinfo {author} {\bibfnamefont {Alberto}\
  \bibnamefont {Crepaldi}}, \bibinfo {author} {\bibfnamefont {Michele}\
  \bibnamefont {Zacchigna}}, \bibinfo {author} {\bibfnamefont {Cephise}\
  \bibnamefont {Cacho}}, \bibinfo {author} {\bibfnamefont {I.~C.~Edmond}\
  \bibnamefont {Turcu}}, \bibinfo {author} {\bibfnamefont {Emma}\ \bibnamefont
  {Springate}}, \bibinfo {author} {\bibfnamefont {Felix}\ \bibnamefont
  {Fromm}}, \bibinfo {author} {\bibfnamefont {Christian}\ \bibnamefont
  {Raidel}}, \bibinfo {author} {\bibfnamefont {Thomas}\ \bibnamefont
  {Seyller}}, \bibinfo {author} {\bibfnamefont {Fulvio}\ \bibnamefont
  {Parmigiani}}, \bibinfo {author} {\bibfnamefont {Marco}\ \bibnamefont
  {Grioni}}, \ and\ \bibinfo {author} {\bibfnamefont {Philip}\ \bibnamefont
  {Hofmann}},\ }\bibfield  {title} {\enquote {\bibinfo {title} {Direct view of
  hot carrier dynamics in graphene},}\ }\href {\doibase
  10.1103/PhysRevLett.111.027403} {\bibfield  {journal} {\bibinfo  {journal}
  {Phys. Rev. Lett.}\ }\textbf {\bibinfo {volume} {111}},\ \bibinfo {pages}
  {027403} (\bibinfo {year} {2013})}\BibitemShut {NoStop}%
\bibitem [{\citenamefont {Boyd}(2008)}]{Boyd}%
  \BibitemOpen
  \bibfield  {author} {\bibinfo {author} {\bibfnamefont {R.W.}\ \bibnamefont
  {Boyd}},\ }\href {https://books.google.co.in/books?id=uoRUi1Yb7ooC} {\emph
  {\bibinfo {title} {Nonlinear Optics}}},\ Nonlinear Optics Series\ (\bibinfo
  {publisher} {Elsevier Science},\ \bibinfo {year} {2008})\BibitemShut
  {NoStop}%
\bibitem [{\citenamefont {Schubert}\ \emph {et~al.}(2014)\citenamefont
  {Schubert}, \citenamefont {Hohenleutner}, \citenamefont {Langer},
  \citenamefont {Urbanek}, \citenamefont {Lange}, \citenamefont {Huttner},
  \citenamefont {Golde}, \citenamefont {Meier}, \citenamefont {Kira},
  \citenamefont {Koch},\ and\ \citenamefont {Huber}}]{Schubert2014}%
  \BibitemOpen
  \bibfield  {author} {\bibinfo {author} {\bibfnamefont {O.}~\bibnamefont
  {Schubert}}, \bibinfo {author} {\bibfnamefont {M.}~\bibnamefont
  {Hohenleutner}}, \bibinfo {author} {\bibfnamefont {F.}~\bibnamefont
  {Langer}}, \bibinfo {author} {\bibfnamefont {B.}~\bibnamefont {Urbanek}},
  \bibinfo {author} {\bibfnamefont {C.}~\bibnamefont {Lange}}, \bibinfo
  {author} {\bibfnamefont {U.}~\bibnamefont {Huttner}}, \bibinfo {author}
  {\bibfnamefont {D.}~\bibnamefont {Golde}}, \bibinfo {author} {\bibfnamefont
  {T.}~\bibnamefont {Meier}}, \bibinfo {author} {\bibfnamefont
  {M.}~\bibnamefont {Kira}}, \bibinfo {author} {\bibfnamefont {S.~W.}\
  \bibnamefont {Koch}}, \ and\ \bibinfo {author} {\bibfnamefont
  {R.}~\bibnamefont {Huber}},\ }\bibfield  {title} {\enquote {\bibinfo {title}
  {{Sub-cycle control of terahertz high-harmonic generation by dynamical Bloch
  oscillations}},}\ }\href {\doibase 10.1038/nphoton.2013.349} {\bibfield
  {journal} {\bibinfo  {journal} {Nature Photonics}\ }\textbf {\bibinfo
  {volume} {8}},\ \bibinfo {pages} {119--123} (\bibinfo {year}
  {2014})}\BibitemShut {NoStop}%
\bibitem [{\citenamefont {Zhang}\ and\ \citenamefont {Voss}(2011)}]{Zhang1}%
  \BibitemOpen
  \bibfield  {author} {\bibinfo {author} {\bibfnamefont {Zheshen}\ \bibnamefont
  {Zhang}}\ and\ \bibinfo {author} {\bibfnamefont {Paul~L.}\ \bibnamefont
  {Voss}},\ }\bibfield  {title} {\enquote {\bibinfo {title} {Full-band
  quantum-dynamical theory of saturation and four-wave mixing in graphene},}\
  }\href {\doibase 10.1364/OL.36.004569} {\bibfield  {journal} {\bibinfo
  {journal} {Opt. Lett.}\ }\textbf {\bibinfo {volume} {36}},\ \bibinfo {pages}
  {4569--4571} (\bibinfo {year} {2011})}\BibitemShut {NoStop}%
\bibitem [{\citenamefont {Burghoff}\ \emph {et~al.}(2014)\citenamefont
  {Burghoff}, \citenamefont {Kao}, \citenamefont {Han}, \citenamefont {Chan},
  \citenamefont {Cai}, \citenamefont {Yang}, \citenamefont {Hayton},
  \citenamefont {Gao}, \citenamefont {Reno},\ and\ \citenamefont
  {Hu}}]{Burghoff2014}%
  \BibitemOpen
  \bibfield  {author} {\bibinfo {author} {\bibfnamefont {David}\ \bibnamefont
  {Burghoff}}, \bibinfo {author} {\bibfnamefont {Tsung-Yu}\ \bibnamefont
  {Kao}}, \bibinfo {author} {\bibfnamefont {Ningren}\ \bibnamefont {Han}},
  \bibinfo {author} {\bibfnamefont {Chun Wang~Ivan}\ \bibnamefont {Chan}},
  \bibinfo {author} {\bibfnamefont {Xiaowei}\ \bibnamefont {Cai}}, \bibinfo
  {author} {\bibfnamefont {Yang}\ \bibnamefont {Yang}}, \bibinfo {author}
  {\bibfnamefont {Darren~J.}\ \bibnamefont {Hayton}}, \bibinfo {author}
  {\bibfnamefont {Jian-Rong}\ \bibnamefont {Gao}}, \bibinfo {author}
  {\bibfnamefont {John~L.}\ \bibnamefont {Reno}}, \ and\ \bibinfo {author}
  {\bibfnamefont {Qing}\ \bibnamefont {Hu}},\ }\bibfield  {title} {\enquote
  {\bibinfo {title} {{Terahertz laser frequency combs}},}\ }\href {\doibase
  10.1038/nphoton.2014.85} {\bibfield  {journal} {\bibinfo  {journal} {Nature
  Photonics}\ }\textbf {\bibinfo {volume} {8}},\ \bibinfo {pages} {462--467}
  (\bibinfo {year} {2014})}\BibitemShut {NoStop}%
\bibitem [{\citenamefont {Aversa}\ and\ \citenamefont {Sipe}(1995)}]{Aversa}%
  \BibitemOpen
  \bibfield  {author} {\bibinfo {author} {\bibfnamefont {Claudio}\ \bibnamefont
  {Aversa}}\ and\ \bibinfo {author} {\bibfnamefont {J.~E.}\ \bibnamefont
  {Sipe}},\ }\bibfield  {title} {\enquote {\bibinfo {title} {Nonlinear optical
  susceptibilities of semiconductors: Results with a length-gauge analysis},}\
  }\href {\doibase 10.1103/PhysRevB.52.14636} {\bibfield  {journal} {\bibinfo
  {journal} {Phys. Rev. B}\ }\textbf {\bibinfo {volume} {52}},\ \bibinfo
  {pages} {14636--14645} (\bibinfo {year} {1995})}\BibitemShut {NoStop}%
\bibitem [{\citenamefont {Singh}\ \emph {et~al.}(2017)\citenamefont {Singh},
  \citenamefont {Bolotin}, \citenamefont {Ghosh},\ and\ \citenamefont
  {Agarwal}}]{PhysRevB.95.155421}%
  \BibitemOpen
  \bibfield  {author} {\bibinfo {author} {\bibfnamefont {Ashutosh}\
  \bibnamefont {Singh}}, \bibinfo {author} {\bibfnamefont {Kirill~I.}\
  \bibnamefont {Bolotin}}, \bibinfo {author} {\bibfnamefont {Saikat}\
  \bibnamefont {Ghosh}}, \ and\ \bibinfo {author} {\bibfnamefont {Amit}\
  \bibnamefont {Agarwal}},\ }\bibfield  {title} {\enquote {\bibinfo {title}
  {Nonlinear optical conductivity of a generic two-band system with application
  to doped and gapped graphene},}\ }\href {\doibase 10.1103/PhysRevB.95.155421}
  {\bibfield  {journal} {\bibinfo  {journal} {Phys. Rev. B}\ }\textbf {\bibinfo
  {volume} {95}},\ \bibinfo {pages} {155421} (\bibinfo {year}
  {2017})}\BibitemShut {NoStop}%
\bibitem [{\citenamefont {Singh}\ \emph
  {et~al.}(2018{\natexlab{a}})\citenamefont {Singh}, \citenamefont {Ghosh},\
  and\ \citenamefont {Agarwal}}]{PhysRevB.97.045402}%
  \BibitemOpen
  \bibfield  {author} {\bibinfo {author} {\bibfnamefont {Ashutosh}\
  \bibnamefont {Singh}}, \bibinfo {author} {\bibfnamefont {Saikat}\
  \bibnamefont {Ghosh}}, \ and\ \bibinfo {author} {\bibfnamefont {Amit}\
  \bibnamefont {Agarwal}},\ }\bibfield  {title} {\enquote {\bibinfo {title}
  {Nonlinear, anisotropic, and giant photoconductivity in intrinsic and doped
  graphene},}\ }\href {\doibase 10.1103/PhysRevB.97.045402} {\bibfield
  {journal} {\bibinfo  {journal} {Phys. Rev. B}\ }\textbf {\bibinfo {volume}
  {97}},\ \bibinfo {pages} {045402} (\bibinfo {year}
  {2018}{\natexlab{a}})}\BibitemShut {NoStop}%
\bibitem [{\citenamefont {Singh}\ \emph
  {et~al.}(2018{\natexlab{b}})\citenamefont {Singh}, \citenamefont {Ghosh},\
  and\ \citenamefont {Agarwal}}]{PhysRevB.97.205420}%
  \BibitemOpen
  \bibfield  {author} {\bibinfo {author} {\bibfnamefont {Ashutosh}\
  \bibnamefont {Singh}}, \bibinfo {author} {\bibfnamefont {Saikat}\
  \bibnamefont {Ghosh}}, \ and\ \bibinfo {author} {\bibfnamefont {Amit}\
  \bibnamefont {Agarwal}},\ }\bibfield  {title} {\enquote {\bibinfo {title}
  {Nonlinear and anisotropic polarization rotation in two-dimensional dirac
  materials},}\ }\href {\doibase 10.1103/PhysRevB.97.205420} {\bibfield
  {journal} {\bibinfo  {journal} {Phys. Rev. B}\ }\textbf {\bibinfo {volume}
  {97}},\ \bibinfo {pages} {205420} (\bibinfo {year}
  {2018}{\natexlab{b}})}\BibitemShut {NoStop}%
\bibitem [{\citenamefont {Chaves}\ \emph {et~al.}(2016)\citenamefont {Chaves},
  \citenamefont {Peres},\ and\ \citenamefont {Low}}]{PhysRevB.94.195438}%
  \BibitemOpen
  \bibfield  {author} {\bibinfo {author} {\bibfnamefont {A.~J.}\ \bibnamefont
  {Chaves}}, \bibinfo {author} {\bibfnamefont {N.~M.~R.}\ \bibnamefont
  {Peres}}, \ and\ \bibinfo {author} {\bibfnamefont {Tony}\ \bibnamefont
  {Low}},\ }\bibfield  {title} {\enquote {\bibinfo {title} {Pumping electrons
  in graphene to the $m$ point in the brillouin zone: Emergence of anisotropic
  plasmons},}\ }\href {\doibase 10.1103/PhysRevB.94.195438} {\bibfield
  {journal} {\bibinfo  {journal} {Phys. Rev. B}\ }\textbf {\bibinfo {volume}
  {94}},\ \bibinfo {pages} {195438} (\bibinfo {year} {2016})}\BibitemShut
  {NoStop}%
\bibitem [{Note1()}]{Note1}%
  \BibitemOpen
  \bibinfo {note} {Strictly speaking, there is an additional term of the form
  $i (\Omega _{\protect \boldsymbol k}^{cc} -\Omega _{\protect \boldsymbol
  k}^{vv}) p_{\protect \boldsymbol k}$ on the right hand side of Eq.~\protect
  \textup {\hbox {\mathsurround \z@ \protect \normalfont (\ignorespaces \ref
  {OBE_p}\unskip \@@italiccorr )}}. However this term can be safely neglected
  as it 1) oscillates with a frequency $~ 2\omega _p$ \cite
  {PhysRevB.95.155421} and 2) its impact on the final steady state dynamics
  turns out to be very small \cite {PhysRevB.97.205420}.}\BibitemShut {Stop}%
\bibitem [{\citenamefont {{Semnani}}\ \emph {et~al.}(2018)\citenamefont
  {{Semnani}}, \citenamefont {{Jago}}, \citenamefont {{Safavi-Naeini}},
  \citenamefont {{Hamed Majedi}}, \citenamefont {{Malic}},\ and\ \citenamefont
  {{Tassin}}}]{2018arXiv180610123S}%
  \BibitemOpen
  \bibfield  {author} {\bibinfo {author} {\bibfnamefont {Behrooz}\ \bibnamefont
  {{Semnani}}}, \bibinfo {author} {\bibfnamefont {Roland}\ \bibnamefont
  {{Jago}}}, \bibinfo {author} {\bibfnamefont {Safieddin}\ \bibnamefont
  {{Safavi-Naeini}}}, \bibinfo {author} {\bibfnamefont {Amir}\ \bibnamefont
  {{Hamed Majedi}}}, \bibinfo {author} {\bibfnamefont {Ermin}\ \bibnamefont
  {{Malic}}}, \ and\ \bibinfo {author} {\bibfnamefont {Philippe}\ \bibnamefont
  {{Tassin}}},\ }\bibfield  {title} {\enquote {\bibinfo {title} {{Anomalous
  optical saturation of low-energy Dirac states in graphene and its implication
  for nonlinear optics}},}\ }\href@noop {} {\bibfield  {journal} {\bibinfo
  {journal} {arXiv e-prints}\ ,\ \bibinfo {eid} {arXiv:1806.10123}} (\bibinfo
  {year} {2018})},\ \Eprint {http://arxiv.org/abs/1806.10123} {arXiv:1806.10123
  [cond-mat.mes-hall]} \BibitemShut {NoStop}%
\bibitem [{\citenamefont {Yoshino}(2013)}]{Yoshino:13}%
  \BibitemOpen
  \bibfield  {author} {\bibinfo {author} {\bibfnamefont {Toshihiko}\
  \bibnamefont {Yoshino}},\ }\bibfield  {title} {\enquote {\bibinfo {title}
  {Theory for oblique-incidence magneto-optical faraday and kerr effects in
  interfaced monolayer graphene and their characteristic features},}\ }\href
  {\doibase 10.1364/JOSAB.30.001085} {\bibfield  {journal} {\bibinfo  {journal}
  {J. Opt. Soc. Am. B}\ }\textbf {\bibinfo {volume} {30}},\ \bibinfo {pages}
  {1085--1091} (\bibinfo {year} {2013})}\BibitemShut {NoStop}%
\bibitem [{Note2()}]{Note2}%
  \BibitemOpen
  \bibinfo {note} {The simplified expressions in the last column are obtained
  using $\alpha _F\ll 1$, along with $\sigma ^2_{xy} \ll \sigma _{0}$ -- which
  works in the case of graphene.}\BibitemShut {Stop}%
\bibitem [{\citenamefont {Nair}\ \emph {et~al.}(2008)\citenamefont {Nair},
  \citenamefont {Blake}, \citenamefont {Grigorenko}, \citenamefont {Novoselov},
  \citenamefont {Booth}, \citenamefont {Stauber}, \citenamefont {Peres},\ and\
  \citenamefont {Geim}}]{Nair2008}%
  \BibitemOpen
  \bibfield  {author} {\bibinfo {author} {\bibfnamefont {R.~R.}\ \bibnamefont
  {Nair}}, \bibinfo {author} {\bibfnamefont {P.}~\bibnamefont {Blake}},
  \bibinfo {author} {\bibfnamefont {A.~N.}\ \bibnamefont {Grigorenko}},
  \bibinfo {author} {\bibfnamefont {K.~S.}\ \bibnamefont {Novoselov}}, \bibinfo
  {author} {\bibfnamefont {T.~J.}\ \bibnamefont {Booth}}, \bibinfo {author}
  {\bibfnamefont {T.}~\bibnamefont {Stauber}}, \bibinfo {author} {\bibfnamefont
  {N.~M.~R.}\ \bibnamefont {Peres}}, \ and\ \bibinfo {author} {\bibfnamefont
  {A.~K.}\ \bibnamefont {Geim}},\ }\bibfield  {title} {\enquote {\bibinfo
  {title} {Fine structure constant defines visual transparency of graphene},}\
  }\href {\doibase 10.1126/science.1156965} {\bibfield  {journal} {\bibinfo
  {journal} {Science}\ }\textbf {\bibinfo {volume} {320}},\ \bibinfo {pages}
  {1308} (\bibinfo {year} {2008})}\BibitemShut {NoStop}%
\bibitem [{\citenamefont {Mak}\ \emph {et~al.}(2008)\citenamefont {Mak},
  \citenamefont {Sfeir}, \citenamefont {Wu}, \citenamefont {Lui}, \citenamefont
  {Misewich},\ and\ \citenamefont {Heinz}}]{Mak}%
  \BibitemOpen
  \bibfield  {author} {\bibinfo {author} {\bibfnamefont {Kin~Fai}\ \bibnamefont
  {Mak}}, \bibinfo {author} {\bibfnamefont {Matthew~Y.}\ \bibnamefont {Sfeir}},
  \bibinfo {author} {\bibfnamefont {Yang}\ \bibnamefont {Wu}}, \bibinfo
  {author} {\bibfnamefont {Chun~Hung}\ \bibnamefont {Lui}}, \bibinfo {author}
  {\bibfnamefont {James~A.}\ \bibnamefont {Misewich}}, \ and\ \bibinfo {author}
  {\bibfnamefont {Tony~F.}\ \bibnamefont {Heinz}},\ }\bibfield  {title}
  {\enquote {\bibinfo {title} {Measurement of the optical conductivity of
  graphene},}\ }\href {\doibase 10.1103/PhysRevLett.101.196405} {\bibfield
  {journal} {\bibinfo  {journal} {Phys. Rev. Lett.}\ }\textbf {\bibinfo
  {volume} {101}},\ \bibinfo {pages} {196405} (\bibinfo {year}
  {2008})}\BibitemShut {NoStop}%
\bibitem [{Note3()}]{Note3}%
  \BibitemOpen
  \bibinfo {note} {Reflectivity and polarization rotation for the newly
  generated sideband are defined in reference to the amplitude of backward
  propagating field $r^{(-1)}$, and its polarization relative to the
  polarization of the incident pump beam.}\BibitemShut {Stop}%
\bibitem [{\citenamefont {Katsnelson}(2012)}]{katsnelson2012graphene}%
  \BibitemOpen
  \bibfield  {author} {\bibinfo {author} {\bibfnamefont {M.~I.}\ \bibnamefont
  {Katsnelson}},\ }\href {https://books.google.co.in/books?id=FgwW3aVRBT0C}
  {\emph {\bibinfo {title} {Graphene: Carbon in Two Dimensions}}}\ (\bibinfo
  {publisher} {Cambridge University Press},\ \bibinfo {year}
  {2012})\BibitemShut {NoStop}%
\end{thebibliography}%
\end{document}